\documentclass[12pt]{iopart}
\usepackage{graphicx}
\usepackage{color}

\newcommand {\beq} {\begin{equation}}
\newcommand {\eeq} {\end{equation}}
 \newcommand {\ber}{\begin{eqnarray*}}
 \newcommand {\eer} {\end{eqnarray*}}
\newcommand {\bea}{\begin{eqnarray}}
 \newcommand {\eea} {\end{eqnarray}}
\newcommand {\beqa}{\begin{eqnarray}}
 \newcommand {\eeqa} {\end{eqnarray}}
\newcommand{\be}{\begin{equation}}
\newcommand{\ee}{\end{equation}}
\newcommand{\eq}[1]{(\ref{#1})}

\def\nn{\nonumber}

\def\rr{{\rm r}} \def\rs{{\rm s}}\def\ri{{\rm i}}
\def\ra{{\rm a}} \def\rb{{\rm b}}
\def\del{\partial}

\newcommand{\bra}[1]{\langle #1|}
\newcommand{\ket}[1]{|#1 \rangle}

\def\rr{{\rm r}} \def\rs{{\rm s}}
\def\del{\partial}

\def\a{\alpha}          
\def\b{\beta}           
    
\def\d{\delta}    

\def\l{\lambda} \def\L{\Lambda}
\def\m{\mu}     \def\n{\nu} \def\o{\omega}
     \def\P{\Pi}

  \def\S{\Sigma}

\def\th{\theta}

\font\mybb=msbm10 at 12pt

\def\bb#1{\hbox{\mybb#1}}

\def\zet{{\bb{Z}}}

\def\real{{\bb{R}}}
\newcommand{\ex}[1]{{\rm e}^{#1}} \def\ii{{\rm i}}

\def\P{{\bb{P}}}
 
\def\cN{{\cal N}} 
\def\del{\partial}
\def\ii{{\rm i}}

\def\rr{{\rm r}}

\newcommand{\Ttr}{{\rm Tr}}

\begin{document}

\hfill{CERN-PH-TH/2004-008}

\hfill{QMUL-PH-04-01}

\hfill{LPTHE-P04-01}

\title[The PP-wave/CFT duality: a Status Report]{The Duality between IIB 
String Theory on PP-wave and ${\cal N}=4$ SYM: a Status Report}

\author{Rodolfo Russo \footnote{On leave of absence from {\em Queen Mary, University of London},  E1 4NS London, UK.\\ Email: {\tt Rodolfo.Russo@cern.ch}}
}

\address{CERN, Geneva 23, CH-1211 Switzerland.}

\author{Alessandro Tanzini \footnote{Email: {\tt tanzini@lpthe.jussieu.fr}}
}
\setcounter{footnote}{0}
\address{Laboratoire de Physique Th\'eorique et Hautes Energies,
Universit\'e de Paris VI, 4 Place Jussieu, case 126, 75252 Paris CEDEX 05,
France}

\begin{abstract}

The aim of this report is to give an overview of the duality between
type IIB string theory on the maximally supersymmetric PP-wave and the
BMN sector of ${\cal N}=4$ Super Yang-Mills theory. The general
features of the string and the field theory descriptions are 
reviewed, but the main focus of this report is on the comparison
between the two sides of the duality. In particular, it is first
explained how free IIB strings emerge on the gauge theory side and then
the generalizations of this relation to the full interacting theory
are considered. An ``historical'' approach is taken and the various
proposals presented in the literature are described. 

\end{abstract}

%Uncomment for PACS numbers title message
%\pacs{00.00, 20.00, 42.10}

% Uncomment for Submitted to journal title message
%\submitto{\JPA}

% Comment out if separate title page not required
% \maketitle

\section{Introduction}

In~\cite{Berenstein:2002jq}, Berenstein, Maldacena and Nastase
proposed a very concrete relation between type IIB string theory on
the maximally supersymmetric PP-wave and a particular subsector of the
${\cal N}=4$ Super Yang-Mills theory. This proposal immediately
attracted a lot of attention and triggered a great deal of activity
both in the analysis of the original physical situation and in the
extension of these ideas to other interesting cases. The reason for
this intense activity is twofold. On the one hand, PP-waves are
interesting in themselves, since they provide a perfect arena for
studying string theory in backgrounds that are very different from
Minkowski flat space. For instance, PP-waves can be used to test our
understanding of the string dynamics in presence of non-trivial
Ramond-Ramond fields and provide {\em tractable} backgrounds that are
nevertheless curved and not even asymptotically flat. On the other
hand the relation proposed in~\cite{Berenstein:2002jq} is the first
example of a direct connection between a four dimensional gauge theory
and a string theory that can be quantized. In particular, the relation
between strings on the maximally supersymmetric PP-wave and the ${\cal
  N}=4$ Super Yang-Mills theory appears to be a ``corollary'' of the
usual AdS/CFT duality~\cite{Maldacena:1997re}. Thus, at the moment,
the PP-wave background provides the best setup for testing in a
concrete example the various ideas about the gauge/string theory
duality that have been developed since 't~Hooft's seminal
paper~\cite{'tHooft:1974jz}.

Besides the various applications, the study of string dynamics on
PP-wave has a beauty in itself because it connects many ideas coming
from different areas of theoretical physics, like general relativity,
string perturbation theory, gauge theories, integrable systems and
others. This makes the subject quite rich and clearly it is not
possible to summarize all the developments in a single paper.
Fortunately various reviews are already available, each of them
analyzing some particular aspect of the problem. The original proposal
is summarized in~\cite{Maldacena:2003nj}.
Ref.~\cite{Pankiewicz:2003pg} reviews the techniques usually employed
to describe the string dynamics, while~\cite{Plefka:2003nb} focuses
more on the field theory side of the correspondence. The most recent
reviews are~\cite{Sadri:2003pr}, where the PP--wave/SYM duality
proposed in~\cite{Berenstein:2002jq} is described in general,
and~\cite{Tseytlin:2003ii}, where the semiclassical interpretation of
the PP-wave limit and the latest developments in this area are
summarized. For a systematic introduction to the various aspects of
the PP--wave/SYM duality, the reader is referred to the reviews cited
above. In fact, this report does {\em not} provide a detailed
description of the subject, since, for instance, the explicit
derivation of many technical results will be simply omitted. On the
contrary, the aim of this work is to focus on the comparison between
the string and the field theory side of the PP--wave/SYM duality, and
to summarize all the proposals that appeared in the literature on this
issue. Particular attention is devoted to see how these proposals can
be interpreted in the broader context of the string/gauge theory
correspondence {\em \`a la} 't~Hooft.  In this respect it is very
important to stress the connections between the setup presented
in~\cite{Berenstein:2002jq} and the AdS/CFT duality involving the full
${\cal N}=4$ Super Yang-Mills theory originally proposed
in~\cite{Maldacena:1997re}.  The main goal is clearly to use string
theory on PP-waves to learn more about the general properties of the
AdS/CFT duality and to derive results on quantities not protected by
supersymmetry.

The structure of this report is the following. In Section~\ref{basic}
the basic ingredients of the PP-wave/SYM duality are introduced: the
plane-wave solution itself, the dynamics of {\em free} IIB strings on
this background and finally the BMN
dictionary~\cite{Berenstein:2002jq} between the Fock space of the
string states and a subset of gauge invariant composite operators on
the SYM side. This relation between the physical spectra of the two
descriptions can be explained in different ways. At the planar/free
level all these approaches lead to equivalent results, but they
inspired different proposals when the correspondence is generalized to
the full interacting theory. These developments are discussed in
Section~\ref{further}: in particular, on the string side, the vertices
describing the $3$-string interaction are discussed, while on the
gauge theory side the origin of the operator mixing is explained. At
the light of these results, the BMN dictionary between string states
and field theory operators is re-discussed.
In Section~\ref{comparing} the analysis is extended beyond the map
between spectra and the various proposals dealing with dynamical
quantities of the interacting theory are described. Particular
attention is devoted to the approaches of~\cite{Constable:2002hw} and
of~\cite{Gross:2002mh}. 
In the final Section we briefly describe some possible developments in
the field and summarize the open problems both at the conceptual and
at the technical level.

\section{The basic concepts of the duality}\label{basic}

The starting point of the BMN proposal~\cite{Berenstein:2002jq}
is the AdS/CFT duality.  In its strong version this duality states
that the ${\cal N}=4$ $SU(N)$ Super Yang-Mills theory and type IIB
string theory on $AdS_5 \times S^5$ with $N$ units of five form flux
are exactly equivalent.  This conjecture has been extensively studied
in the last six years and, even if a real proof is still far away, a
large amount of evidence has emerged. There are many excellent reviews
about this
subject~\cite{Petersen:1999zh,Aharony:1999ti,DiVecchia:1999yr,Klebanov:2000me,D'Hoker:2002aw},
so there is no need to enter into a detailed description.  Only few
basic facts will instead be reported here:

1) One of the most important features of the AdS/CFT correspondence is
the fact that both descriptions possess the same (super)symmetry
group, whose bosonic part is $SO(4,2)\times SO(6)$. On the string
theory side, this corresponds to the isometry group of the
$AdS_5\times S^5$ background, while in the gauge theory description it
represents the conformal group times the internal $R$--symmetry group
$SO(6)\sim SU(4)$ which rotates the four supersymmetry charges.
Notice that both these groups are {\em exact} symmetries of the full
interacting quantum field theory.
This has far reaching implications both for the duality and for
the study of the quantum effects.

2) Type IIB string theory on the $AdS_5\times S^5$ background is
characterised by three independent parameters: the $N$ units of
five-form fluxes on $S^5$, the string coupling $g_s$, the constant
vacuum expectation value of the axion $\chi_0$. The common radius of
$AdS_5$ and $S^5$ measured in string units is $R^2/\alpha'=\sqrt{4\pi
g_s N}$. On the other hand the parameters of the $\cN=4$ SYM theory
are the rank of the gauge group $N$, the coupling $g_{\rm YM}$, from
which one can define the 't~Hooft coupling $\lambda=g_{\rm YM}^2 N$,
and the vacuum angle
$\theta$.  The dictionary between gauge and string theory parameters
identifies on the one side the five--form flux and the rank of the
gauge group, and on the other the coupling constants\footnote{The
$U(N)$ matrices are normalized as
$\Ttr({T^aT^b})=\frac{1}{2}\delta^{ab}$. If one chooses a different
normalization, then Eqs.\eq{coupl} and \eq{radius} are modified. In
general one has $g_{YM}^2 = 2 r \pi g_s$, where $r$ is related to the
gauge group Casimir: $\Ttr({T^aT^b})=\frac{1}{r}\delta^{ab}$.} 
\be 
\tau_{\rm YM}=\frac{\theta}{2\pi}+ i\frac{4\pi}{g_{\rm
YM}^2}=\chi_0+\frac{i}{g_s}~.
\label{coupl}
\ee
Notice that this implies a relation between the $AdS_5\times S^5$
radius and the 't~Hooft coupling
\be
\frac{R^2}{\alpha'}=\sqrt{\lambda} \ \ .
\label{radius}
\ee
This relation has an important consequence: the regime where the
supergravity description of the $AdS_5\times S^5$ dynamics is reliable
$R^2 \gg \a'$, corresponds to a strongly coupled gauge theory
$\lambda \gg 1$. On the other hand, when the perturbative expansion of
gauge theory is reliable, the radius of the corresponding bulk
geometry is small and the full string dynamics is needed in the dual
description.  For this reason, most of the explicit checks of
the AdS/CFT duality involved protected quantities, {\em i .e. }
particular observables that do not receive quantum corrections in
$\lambda$. Clearly in this case the supergravity computation in the
bulk has to match directly the perturbative SYM result. A notable
exception to this limitation is represented by the study of the Wilson
loops, see~\cite{Semenoff:2002kk} and references therein for a review.

3) Since the string and the gauge theory descriptions are supposed to
be equivalent, there must be an isomorphism between the Hilbert spaces
representing the spectra of the two theories. This isomorphism is
largely unknown and actually even the spectrum itself is not fully
understood. However many properties of this mapping have been derived
in the last years. For instance we know that a special role is played
by the single trace operators since they are related to single
particle (string) states on the AdS side, unless one is dealing with
very ``big'' operators (we shall return to this point later in
Section~\ref{bmnrev}). Multiple particle states simply correspond in
the gauge theory to products of non--coincident single trace operators
separately normal ordered, while real multi--trace operators, {\em
  i.e.} operators made out of many traces with fields evaluated in the
same point and globally normal ordered, should correspond to {\em
  bound states} of the elementary string excitations. Of course, the
AdS/CFT dictionary should map not only the various states of the two
spectra, but also their quantum numbers. In particular, the energy of
the string states is directly related to the conformal dimension of
the corresponding gauge theory
operators~\cite{Gubser:1998bc,Witten:1998qj}. At the level of
supergravity excitations this prediction can be tested explicitly:
from the quadratic part of the effective action one can extract the
mass of the various supergravity fields which is directly related to
the conformal dimensions in gauge theory, see for instance Eq.(5.21)
of~\cite{D'Hoker:2002aw}.

4) The correspondence between gauge and string theory goes beyond the
level of the free theory. In fact, it is natural to think that the
isomorphism between the spectra implies also the coincidence of the
correlation functions between string states and gauge theory
operators. This idea has been made precise
in~\cite{Gubser:1998bc,Witten:1998qj}, where it has been proposed that
the string partition function on $AdS_5\times S^5$, subject to
particular boundary conditions at the conformal boundary, is equal to
the generator of the field theory correlators.

\subsection{The maximally supersymmetric PP-wave} \label{maximally}

The gravitational wave relevant to the following analysis
is~\cite{Blau:2001ne} 
\bea \label{pwave-back}
&& g_{-+} = g_{+-} = -2,\quad g_{++} = - \mu^2 \sum_{I=1}^8 x_I x^I,
\quad g_{IJ} = 
\delta_{IJ}~,~ I,J=1,\ldots,8~~,   \nonumber \\ 
&&F_{+1234} = F_{+5678} = 2\mu~,~~ \phi = {\rm constant}~, 
\eea
where $F$ is the R--R five form and $\phi$ the dilaton field.
It is not difficult to see that this field configuration solves the
equations of motion of type IIB supergravity,
but what makes this background interesting is its high degree of
symmetry. It possesses $14$ obvious bosonic symmetries: the shifts of
the light-cone coordinates $x^\pm \to x^\pm + c$ and the separate
rotations of the two groups of directions $(1234)$ and $(5678)$. In
fact the $5$-form $F$ breaks the $SO(8)$ symmetry of the metric down
to $SO(4) \times SO(4)$.  It is worth to remark that the PP--wave
solution Eq.\eq{pwave-back} displays a discrete symmetry $\zet_2$
exchanging the two $SO(4)$ groups. The shifts in the transverse
coordinates $x^I$ are broken by the $g_{++}$ element of the
metric. They are substituted by $x^+$--dependent transformations
${\cal P}^I$. Moreover, the metric in Eq.\eq{pwave-back} in invariant
under the rotations in the $(+,I)$ directions $J^{+I}$. Summarizing,
the gravitational wave in Eq.\eq{pwave-back} displays $30$ bosonic
symmetries, exactly as the $AdS_5\times S^5$ background.  Also the
number of fermionic symmetries is the same, and it is given by the
$32$ fermionic charges $Q^+,\bar Q^+$ and $Q^-,\bar Q^-$. This makes
the PP--wave background particularly interesting, since it is a new
maximally supersymmetric solution with non--trivial curvature. Clearly
this background is closely related to the $AdS_5\times S^5$ geometry
and in fact it can be obtained~\cite{Blau:2002dy} as a particular
limit, the \textit{Penrose limit}~\cite{Penrose,Gueven:2000ru}, from
the $AdS_5\times S^5$ solution
\bea
ds^2 & = & R^2 \left[- \cosh^2 \! r\, d t^2 + 
d r^2 + \sinh^2 \! r \, d\Omega_3^2 +
\cos^2 \! \theta\, d\psi^2 + 
d\theta^2 + \sin^2 \!\theta\, d\Omega_3^{'2}\right]
\nonumber \\ \label{ads5s5}
F_5 & = & \frac{1}{R} \;\left(d \, V_{AdS_5} +d \, V_{S^5}\right)~~,~~  
\phi = {\rm constant}~. 
\eea
Let us briefly recall how the two backgrounds Eqs.\eq{ads5s5} and
\eq{pwave-back} are connected. Starting from $AdS_5\times S^5$, we
first introduce the light-cone coordinates
\beq\label{pl}
 x^+ =\frac{t+\psi}{2\mu}~,~~\; x^-= \mu R^2 \,
\frac{t-\psi}{2}~,~~ \hat r= R\, r~,~~ y=R\,\theta~,
\eeq
and then take the $R\to\infty$, by keeping $x^\pm$, $\hat r$ and $y$
fixed. Clearly the non trivial step of the procedure is in the limit
itself. When $R\to \infty$, Eqs.\eq{pl} define the null geodesic
$t=\psi\sim x^+$, $\theta=r = 0$, but also the metric becomes large,
since there is an overall factor of $R^2$. Thanks to these two
properties the limit $R\to \infty$ on the geometry is (always) well
defined~\cite{Penrose} and in our case leads to the first line of
Eq.\eq{pwave-back}. These considerations can be extended to the full
supergravity~\cite{Gueven:2000ru} provided that the various forms,
representing the {\em gauge potentials}, scale appropriately. In our
case $A_4 \sim R^4$ is exactly the right behavior to give a
non--trivial and well defined R--R form also in the large $R$ limit.
Moreover various features of the plane wave solutions obtained in this
way can be derived on general grounds~\cite{Blau:2002mw}, by using
only the properties of the Penrose limit. Here we would just like to
remind that the number of (super)symmetry generators can not decrease
in the limit, but the algebra can be only
deformed~\cite{Hatsuda:2002xp}. This automatically ensures that the
background Eq.\eq{pwave-back} is maximally supersymmetric, since it is
derived from $AdS_5 \times S^5$.

It is also important to clarify the physical meaning of the Penrose
limit just performed. First notice that the change of
coordinates Eq.\eq{pl} put at the ``center'' of the space a particular
null geodesic. Then it is clear that the Penrose limit enlarge a small
neighborhood around this geodesic to be the full space, washing away
all the rest of the original background. Equivalently one can say
that the Penrose limit is a {\em truncation} of the physical spectrum
of the original theory, where one focuses only on the excitations that
are confined to live close to the null geodesic used to define the
limit. In particular, these states describe short strings
($\frac{R^2}{\alpha'}\gg 1$) fast rotating around the equator of the
five--sphere. In fact many of the features of the string states on the
PP--wave background can be derived by studying these rotating strings
on $AdS_5\times S^5$ in the semiclassical
approximation~\cite{Gubser:2002tv}.

\subsection{The free IIB strings on PP-wave}\label{freesec}

%Introduce light-cone gauge
The covariant world--sheet action describing the propagation of a free
type IIB superstring on the PP--wave background has been formulated in
\cite{Metsaev:2001bj,Metsaev:2002re}.  The remarkable feature of this
action is that in the light--cone gauge it reduces to a quadratic
action describing free massive bosons and fermions.  In fact, by
imposing the light--cone gauge conditions
$X^+=\alpha'p^+\tau_a \equiv\alpha\tau_a$ and $\Gamma^+\theta=0$ and then
rescaling $(\tau,\sigma) = |\a' p^+| (\tau_a,\sigma_a)$, one
gets from the covariant action written
in~\cite{Metsaev:2001bj,Metsaev:2002re}
\begin{equation}
S_b = \frac{1}{4 \pi \alpha'} \int d\tau \int\limits_{0}^{2 \pi
|\alpha|} \!\!\! d \sigma \left[ 
(\partial_{\tau} X)^2 - (\partial_{\sigma}
X)^2 - \mu^2 X^2 \right]~,
\label{bos45}
\end{equation}
\begin{equation}
S_f = \frac{1}{4 \pi \alpha'} \int d\tau \!
\int\limits_{0}^{2 \pi |\alpha|} \!\!\! d
\sigma \left\{i e(\alpha) \left[{\bar{\theta}} \partial_{\tau} \theta +
\theta \partial_{\tau} {\bar{\theta}} + \theta \partial_{\sigma} \theta + 
{\bar{\theta}} \partial_{\sigma} {\bar{\theta}} \right] - 2 \mu 
{\bar{\theta}} \Pi \theta \right\}\,. \nn \\
\label{act567}
\end{equation}
Lorentz indices have been suppressed for sake of simplicity;
the 8 bosons $X^I$ and the 8 fermions $\theta^a$ are always
contracted with a Kronecker $\delta$ except for the mass term in $S_f$,
where $\Pi = \sigma_3 \otimes 1_{4\times 4}$ appears. As usual, we
indicate with $\a'$ the Regge slope, while $\a = \a' p^+$ is the
rescaled light-cone momentum
and $e(\a) = 1$ if $\a >0$ and $e(\a) = -1$ if $\a <0$. Moreover, we
take $\mu>0$. 
From Eqs.\eq{bos45} and \eq{act567} it is
straightforward to derive the mode expansions, the commutation
relations and the expressions for the free symmetry generators.

The most general solution of the equations of motion derived
from~\eq{bos45} is given by:
\bea
X (\tau , \sigma ) &=& i \sqrt{\frac{\alpha '}{2}} 
\left[ 
\frac{a_0 {\rm e}^{- i \mu \tau}}{\sqrt{\omega_0}} -
\frac{a_{0}^{\dagger}{\rm e}^{i \mu \tau}}{\sqrt{\omega_0}} + 
\sum_{n=1}^{\infty}
\frac{1}{\sqrt{\omega_n}} \left( 
\hat{a}^{1}_{n} {\rm e}^{-i\frac{\omega_n \tau - n \sigma}{|\alpha|}}  
- \hat{a}_{n}^{1\dagger} {\rm e}^{i\frac{\omega_n \tau - n
    \sigma}{|\alpha|}} \right) 
\right. \nn\\ && \left. 
+ \sum_{n=1}^{\infty} \frac{1}{\sqrt{\omega_n}}
\left( \hat{a}^{2}_{n} {\rm e}^{-i\frac{\omega_n \tau + n
      \sigma}{|\alpha |}}  - \hat{a}_{n}^{2\dagger} {\rm
    e}^{i\frac{\omega_n \tau + n 
    \sigma}{|\alpha |}} \right) \right]~, 
\label{solut83}
\eea
where $ \omega_n = \sqrt{n^2 + (\alpha \mu)^2}$.
The conjugate momentum is $P = \partial S_b/\partial\dot{X}=
\frac{1}{2 \pi \alpha'} \partial_{\tau} X$ and the light-cone
Hamiltonian \footnote{As usual the canonical Hamiltonian is defined as
  the $\partial_\tau$, while the light cone Hamiltonian is
  identified with the $\partial_{x^+}$ generator.  However the two are
  almost identical, because of the light-cone gauge condition 
  $X^+ = e(\a) \tau$. 
  The only difference is the presence of an additional sign
  $e(\a)$ that is included in~\eq{hamil63}.} is
\begin{equation}
H_b = \frac{1}{\alpha} \left[ \frac{\omega_0}{2} (a_{0}^{\dagger}
a_{0} + a_0 a_{0}^{\dagger} )+ \sum_{i=1}^{2} \sum_{n=1}^{\infty} 
\frac{\omega_{n}}{2} (\hat{a}^{i\dagger}_{n} \hat{a}^{i}_{n} +
\hat{a}_{n}^{i} \hat{a}_{n}^{i \dagger} )\right]~.
\label{hamil63}
\end{equation}
The canonical commutation relations $[ X(\tau, \sigma) , P(\tau,
\sigma') ] = i \delta (\sigma - \sigma' )$ 
are satisfied if the oscillators commute as follows:
\begin{equation}
[a_0 , a_{0}^{\dagger}] =1~~,~~[\hat{a}^{i}_{n} ,
  \hat{a}^{j\dagger}_{m}] = \delta_{nm} \delta^{ij}~,
\label{commu73}
\end{equation}
while all other commutators are vanishing. 
The tower of the string states is built by acting on the string vacuum 
$\ket{0,p^+}$ with the creation operators $\hat a_{n}^{\dagger}\equiv \hat a_{n}^{1\dagger}$
and $\hat a_{-n}^{\dagger}\equiv \hat a_{n}^{2\dagger}$. To describe the 3--string interaction
vertex it is useful to introduce the
following combinations:
\begin{equation}
a_n = \frac{1}{\sqrt{2}} (\hat{a}_{n} + \hat{a}_{-n} )~~,~~
a_{-n} = \frac{i}{\sqrt{2}} (\hat{a}_{n} - \hat{a}_{-n} )
~~,~~~ n\geq 1~.
\label{new76}
\end{equation}
In terms of them we can compute:
\begin{equation}
X (\tau=0 , \sigma) = x_0 + \sqrt{2} \sum_{n=1}^{\infty}
\left[ x_n \cos \frac{n \sigma}{|\alpha|} + x_{-n} \sin \frac{n
\sigma}{|\alpha|} \right]
\label{x78}
\end{equation}
where
\be\label{xxx53}
x_{n} = i
\sqrt{\frac{\alpha'}{2 \omega_n}} (a_n - a_{n}^{\dagger})~
~n\geq 0~~,~~
x_{-n} = i \sqrt{\frac{\alpha'}{2 \omega_n}} 
(a_{-n} - a_{-n}^{\dagger})~~ n<0~,
\ee
and
\begin{equation}
P (\tau=0 , \sigma ) =
\frac{1}{2 \pi |\alpha|} \left[ p_0 + \sqrt{2}\sum_{n=1}^{\infty} \left(
p_n \cos \frac{n \sigma}{|\alpha|} + p_{-n} \sin \frac{n\sigma}{|\alpha|}
  \right) \right]
\label{expa32}
\end{equation}
where
\be\label{ppp73}
p_{n} = i
\sqrt{\frac{\omega_n}{2 \alpha'}} (a_n + a_{n}^{\dagger})~
~n\geq 0~~,~~
p_{-n} = i \sqrt{\frac{\omega_n}{2 \alpha'}} 
(a_{-n} + a_{-n}^{\dagger})~~ n<0~.
\ee
Concerning now the fermions, we introduce the real Grassmann variables
\begin{equation}
\theta = \frac{\theta^1 + i \theta^2}{\sqrt{2}}~~,~~{\bar{\theta}} =
\frac{\theta^1 - i \theta^2}{\sqrt{2}}
\label{the54}
\end{equation}
In term of these variables, the equations of motion are
\begin{equation}
e(\alpha ) \partial_{+} \theta^1 - \mu \Pi \theta^2=0~~,~~
e(\alpha ) \partial_{-} \theta^2 + \mu \Pi \theta^1=0~~,~~
\label{moti976}
\end{equation}
Their most general solution can be written as follows:
\begin{eqnarray}
\sqrt{\frac{|\a|}{\a'}}\,
\theta^1 & = & \frac{1}{\sqrt{2}} %c_0
\left( {\rm e}^{- i \mu \tau} \theta_0 + {\rm e}^{i \mu
\tau} \theta_{0}^{\dagger} \right) + \sum_{n=1}^{\infty} c_n
\left[ {\rm e}^{-i \frac{\omega_n \tau - n \sigma}{|\alpha|} }
\theta_{n}^{1} +  {\rm e}^{i \frac{\omega_n \tau - n \sigma}{|\alpha|} }
(\theta^{1}_{n})^{\dagger} \right] 
\nonumber \\
& + & i \sum_{n=1}^{\infty} c_n \frac{\omega_n - n}{\alpha \mu}
\left[ {\rm e}^{-i \frac{\omega_n \tau + n \sigma}{|\alpha|} }
\Pi \theta_{n}^{2} - {\rm e}^{i \frac{\omega_n \tau + n \sigma}{|\alpha|}}
\Pi (\theta^{2}_{n})^{\dagger}\right]~,
\label{theta1xx}
\end{eqnarray}
and
\begin{eqnarray}
\sqrt{\frac{|\a|}{\a'}}\,
\theta^2 & = & \frac{e(\alpha)}{\sqrt{2}} % e(\alpha) c_0
\left(i {\rm e}^{i \mu \tau} \Pi \theta_{0}^{\dagger} -
i {\rm e}^{- i \mu \tau} \Pi \theta_0 \right)
+ \sum_{n=1}^{\infty} c_n
\left[ {\rm e}^{-i \frac{\omega_n \tau + n \sigma}{|\alpha|} }
\theta_{n}^{2} +  {\rm e}^{i \frac{\omega_n \tau + n \sigma}{|\alpha|} }
(\theta^{2}_{n})^{\dagger} \right] 
%\]
\nonumber \\
& - & i \sum_{n=1}^{\infty} c_n \frac{\omega_n - n}{\alpha \mu}
\left[ {\rm e}^{-i \frac{\omega_n \tau - n \sigma}{|\alpha|} }
\Pi \theta_{n}^{1} -  {\rm e}^{i \frac{\omega_n \tau - n \sigma}{|\alpha|} }
\Pi (\theta^{1}_{n})^{\dagger}\right]
\label{theta2xx}~,
\end{eqnarray}
where the coefficients $c_n$ are defined as 
\begin{equation}
c_n = \frac{1}{\sqrt{1 + \rho_{n}^{2}}}~~,~~ \rho_{n} =
\frac{\omega_n -n}{\mu \alpha} \ \ .
\label{cnrhon}
\end{equation}
From the action in Eq.(\ref{act567}) we can define the conjugate
momenta
$\lambda^i \equiv \frac{\delta L}{\delta \partial_{\tau} \theta^i} =
- \frac{i e(\alpha)}{4 \pi \alpha'} \theta^i$
and we get the Hamiltonian
\be
H_f %&=& \frac{i}{4 \pi \alpha'} \int_{0}^{2 \pi |\alpha|}
%d\sigma \sum_{i=1}^{2} \theta^i \partial_{\tau} \theta^i 
\label{hami96} 
=
\frac{1}{|\alpha|}  %\frac{1}{\alpha'}
\left\{ \sum_{n=1}^{\infty} \sum_{i=1}^{2}
\frac{\omega_n}{2} \left( \theta_{n}^{i \dagger} \theta_{n}^{i} -
\theta_{n}^{i} \theta_{n}^{i \dagger} \right) + \frac{\omega_0}{2}
\left(\theta_{0}^{\dagger} \theta_{0} - \theta_{0}
\theta_{0}^{\dagger}   \right)\right\} \ \ .
\ee
The canonical anticommutation relations
$\{ \theta^i ( \sigma, \tau) , \lambda^j ( \sigma' , \tau) \} =
- {i} \delta^{ij} \delta ( \sigma - \sigma')/2$
imply that
$\{ \theta^i ( \sigma, \tau) , \theta^j ( \sigma' , \tau) \}=
2 \pi \alpha' e(\alpha) \delta^{ij} \delta ( \sigma - \sigma')$, and
are satisfied if
\begin{equation}
\{ \theta_{n}^{i} , \theta^{j \dagger}_{m} \} = e(\a)\,
\delta_{nm} \delta^{ij} \ \ .
\label{commu93}
\end{equation}
The tower of the fermionic string excitations is built by acting on
the string vacuum $\ket{0,p^+}$ with the creation operators $\hat
b_n^{\dagger}\equiv e(\alpha)\, \theta_n^{1 \dagger}$ and $\hat
b_{-n}^{\dagger}\equiv e(\alpha)\,\theta_n^{2 \dagger}$. As for the
bosonic case, in order to describe the 3--string interaction it is
useful to introduce the following combinations:
\bea
\sqrt{2} b_n &=& \hat b_{n} + i \hat b_{-n}~~,~~
\sqrt{2} b_{n}^{\dagger} =
\hat b^{\dagger}_{n} - i \hat b^{\dagger}_{-n}~~,~~n>0
\label{bn} \\
\sqrt{2} b_{-n} &=& \hat b_{n} - i \hat b_{-n}~~,~~
\sqrt{2} b_{-n}^{\dagger} =
\hat b^{\dagger}_{n} + i \hat b^{\dagger}_{-n}~~,~~ n>0
\label{b-n}
\eea
and
\begin{equation}
b_0 = \theta_0~~,~~ e(\alpha)b_{0}^{\dagger} = \theta_{0}^{\dagger}~.
\label{b0}
\end{equation}
From Eq.(\ref{commu93}) the following anticommutation relations are
easily derived:
\begin{equation}
\{ b_n, b_{m}^{\dagger} \} = \delta_{nm}~~,~~ \{ b_n , b_m \} = \{
b_{n}^{\dagger} , b_{m}^{\dagger} \} =0~.
\label{bbdagger}
\end{equation}
In terms of the $b_n$'s the Hamiltonian in Eq.(\ref{hami96}) becomes:
\begin{equation}
H_f = \frac{1}{\alpha} \sum_{n= -\infty}^{\infty}
\frac{\omega_n}{2} (b_{n}^{\dagger} b_n -b_n b_{n}^{\dagger})~~.
\label{hamibb}
\end{equation}
Notice that the zero--point energy of the fermionic oscillators cancels
against the bosonic contribution, as can be seen by the comparison between
Eq.(\ref{hamil63}) and Eq.(\ref{hamibb}), and the full free Hamiltonian
can simply be written as follows
\beq \label{fullha}
{H} =  \frac{1}{\alpha}  \sum_{n =-\infty}^{+ \infty} 
\omega_n \left({\hat{a}}^{\dagger}_{n} {\hat{a}}_n + 
{\hat{b}}_{n}^{\dagger} {\hat{b}}_n\right)~.
\eeq

\subsection{The BMN dictionary}\label{bmndsect}

It was shown that the PP--wave background can be obtained as a Penrose
limit of the $AdS_5\times S^5$ geometry, and that this limit can be
described as a truncation of the full string spectrum to a particular
subsector. It is then natural to expect from the AdS/CFT
correspondence that there is a subsector of the $\cN=4$ SYM theory
which is dual to IIB string theory on the PP--wave. In order to
properly define the gauge invariant operators relevant to this duality
we need to recall the content of the the $\cN=4$ SYM. We adopt the
formulation in terms of ${\cal N}=2$ multiplets since this facilitates
the matching with the string side. In fact, this formulation has the
advantage to realize explicitly the $R$-symmetry subgroup
$SU(2)_V\times SU(2)_H\times U(1)_J\subset SU(4)$, where $SU(2)_V,
SU(2)_H$ are respectively the internal symmetry groups of the ${\cal
  N}=2$ vector multiplet and hypermultiplet. In this way we can
naturally identify the $SU(2)_V\times SU(2)_H$ group with one of the
two sets of $SO(4)$ rotations that preserve the solution
Eq.\eq{pwave-back}. The isometries of the PP--wave background contain
another $SO(4)$ group which is identified with the Euclidean rotations
of the $\cN=4$ gauge theory. Finally the $U(1)_J$ group rotates the
complex scalar field $Z$ of the vector multiplet and corresponds, on
the gravity side, to the shift of $x^-$. Actually this identification
can not hold exactly since $x^-\in \real$, while $U(1)_J$ is a compact
generator. It can be valid only if we focus on the gauge theory
operators that have a large $U(1)_J$ charge $J$ and then take the
limit $J\to \infty$. In fact this truncation of the $\cN=4$ spectrum
corresponds to the gauge theory analogue of the Penrose limit
discussed in Section~\ref{maximally}. This important observation can
be made precise by rewriting, thanks to Eq.\eq{pl}, the generators of
the shifts of $x^+$ (the light--cone Hamiltonian) and of $x^-$ (the
$p^+$ momentum) in terms of $\partial/\partial t$ and
$\partial/\partial\psi$. Then one can translate this result on the
gauge theory side obtaining
\beq\label{rf}
\frac{H}{\mu} = \Delta - J  ~~,~~~~
2 \mu p^{+}  = \frac{\Delta + J}{R^2}~,
\eeq
where, following the usual AdS/CFT intuition, the energy
$-i\,\partial/\partial t$ has been mapped to the conformal dimension
$\Delta$ and the angular momentum $-i\,\partial/\partial\psi$ to a
$U(1)_J$ charge. Notice that the factor of $2$ multiplying $p^+$ in
the second relation is consequence of the value of $g_{-+}$. Thus
the gauge invariant operators we need to consider 
have a parametrically large $J$ charge and conformal dimension
($\Delta, J \sim R^2$), while their difference has to be
finite. Because of this, we can approximate $\Delta+J$ with $2 J$ 
\begin{equation}
2 \mu p^{+} = \frac{\Delta + J}{R^2}  \rightarrow \frac{2 J}{R^2}
= \frac{2J}{\alpha' \sqrt{\lambda}}~,
\label{rela46}
\end{equation}
and find that $p^+$ and the $U(1)_J$ charge are identified, as
anticipated.  Summarizing, the Penrose limit translates on the field
theory side in the double scaling
limit~\cite{Kristjansen:2002bb,Constable:2002hw}
\begin{equation}
\Delta \rightarrow \infty~~,~~J \rightarrow \infty~~,~~N \rightarrow
\infty~~{\rm with}~~~\frac{J}{\sqrt{N}}~,~ \Delta - J~,
 ~g^{2}_{YM}~~{\rm fixed}\;,
\label{bmnlimi85}
\end{equation}
which instructs us to keep on the field theory side only the
gauge--invariant operators containing an infinite number of $Z$
fields.  Notice also that this limit implies that the 't~Hooft
coupling $\lambda$ goes to infinity, which from Eq.\eq{radius}
directly corresponds to the Penrose limit $\frac{R^2}{\alpha'}
\rightarrow \infty$.  Nonetheless, the presence of the large quantum
number $J$ allows for the definition of a new parameter $\lambda'$
which is finite in the double scaling limit Eq.\eq{bmnlimi85}. More
precisely, the string dynamics in the PP--wave background is characterized
by the two independent parameters
\begin{equation}
g_2 \equiv \frac{J^2}{N} = 4\pi g_s (\mu\a' p^+)^2
~~,~~
\lambda' \equiv \frac{1}{(\alpha' \mu p^{+})^2} = \frac{\lambda}{J^2}~.
\label{rela44a}
\end{equation}
The relation involving $\lambda'$ follows from Eq.\eq{rela46} and then
combining this with $g_{YM}^2 = 4\pi g_s$, one obtains the first
relation in Eq.\eq{rela44a}. Notice that the string expression of
$\lambda'$ does not contain any power of the string coupling constant;
thus it can be seen as a {classical} quantity related to the
world--sheet dynamics. According to the usual 't~Hooftian intuition
the string world--sheet dynamics resums the {quantum} perturbative
expansion of the dual gauge theory. On the contrary, the string
expression for $g_2$ contains $g_s$ explicitly and thus has to be
interpreted as a space--time coupling, while on the SYM side $g_2$ is
expressed in terms of {classical} quantities only.  Thus $g_2$
corresponds~\cite{Kristjansen:2002bb,Constable:2002hw} to the genus
expansion parameter for the operators satisfying Eq.\eq{bmnlimi85}: on
the string side this is a {quantum} expansion while, on the field
theory side, it classifies topology of the various diagrams. This
means that the sphere approximation (tree-level) in string theory
resums only the planar contributions on the SYM side, the $1$--loop
string diagrams correspond to the field theory contributions that are
planar if drawn on a {torus} and so on. Thus, as in the case of the
full AdS/CFT duality, also for the BMN subsector one can use {\em
  classical} string theory to derive field theory results valid at the
{\em quantum} level, with the caveat the only {\em planar}
contributions have been considered.
A first simple and very interesting example of such a prediction can
be derived from the first of the equations in~\eq{rf} which connects
the string Hamilitonian~\eq{fullha} to the gauge theory dilatation
operator 
\beq
\Delta - J = \frac{H}{\mu} = \sum_{n =-\infty}^{+ \infty}% N_n
\hat\omega_n \left({\hat{a}}^{\dagger}_{n} {\hat{a}}_n + 
{\hat{b}}_{n}^{\dagger} {\hat{b}}_n\right)\,, ~\mbox{with}~
~~\hat\omega_n = \sqrt{1 + n^2 \lambda'}\label{rela45a}~.
\eeq
This can be seen as an operatorial relation: $H$ acts on the Fock
space built with the string oscillators $\hat{a}^\dagger$ and $\hat{b}^\dagger$ and
gives the tree-level (sphere) contribution to the energy of each
state; $\Delta - J$ acts on the gauge invariant operators
satisfying Eq.\eq{bmnlimi85} and gives the planar contribution to their
conformal dimension (minus their $J$--charge). One can
test Eq.\eq{rela45a} by diagonalizing the operators on both sides of the
relation and then building a dictionary between string and gauge theory
states by identifying the eigenvectors. This approach is reviewed
in~\cite{Plefka:2003nb}, thus let us now recall a different way to
derive the same result. The starting point is to use the mapping
between the symmetries in the two descriptions introduced at the
beginning of this section. Then, as in the AdS/CFT case, one can start
build a dictionary between string and gauge theory spectra by
selecting the states with the same quantum numbers. The simplest
possible string state is the vacuum $\ket{v}$, {\em i.e.} the state
which is annihilated by all destruction operators $\hat{a}_n$,
$\hat{b}_n$ and has minimal light--cone energy. The field theory
operator related to this state is a gauge--invariant operator
containing only the complex fields $Z$, $:\Ttr[Z^J]:$, where 
the quantum numbers $p^+$ and $J$ are connected by the Eq.\eq{rela46},
$J^2 = \lambda (\mu \alpha' p^+)^2$.  The next simplest states are the
half--BPS states constructed, on the string side, by acting on the
vacuum with the $\hat{a}_0^\dagger,\; \hat b_0^\dagger$. These
creation operators transform non-trivially under the isometry group of
the background and it is not difficult to find, on the field theory
side, fields that transform in the same way. These fields are the SYM
analogue of the harmonic oscillators and must have $\Delta-J=1$, since
$[H/\mu,\hat{a}_0^\dagger]=1$.
We can construct the field theory operators corresponding to 
the string states created by $\hat{a}_0^\dagger,\; \hat b_0^\dagger$
by inserting these fields as ``impurities'' in the long string
of $Z$ fields representing the string vacuum.
In particular, the impurities associated to the eight bosonic creation operators
$\hat{a}_0^{\dagger}$ are represented by the covariant derivative
acting on the $Z$ field and by the four scalars in the hypermultiplet
(or by the two other complex scalar fields $(\phi^1, \phi^2)$ in the
$\cN =1$ formulation).
The impurities associated to the eight fermionic creation operators
$\hat{b}_0^{\dagger}$ are associated with the Weyl fermions
$\lambda_{\a}^{u}$ of the vector multiplet (for the $\Pi=1$ chirality)
and $\bar\psi_{\dot\a}^{\dot u}$ of the hypermultiplet (for the
$\Pi=-1$ chirality). Alternatively in the ${\cal N}=1$ language they
can be represented by the gaugino $\lambda_{\alpha}$, the superpartner
$\psi^Z_{\alpha}$ of the $Z$ field and the superpartners
$(\bar\psi^1_{\dot\alpha}, \bar\psi^2_{\dot\alpha})$ of the scalar
fields $(\bar\phi_1, \bar\phi_2)$. The first two impurities correspond
to the insertion of string fermionic oscillators of chirality $\Pi =
1$, while the last two to fermionic oscillators of the opposite
chirality $\Pi = -1$. With this input, one can construct gauge theory
operators that transform exactly as the string state considered by
inserting in the string of $Z$ of the vacuum operator the fields
corresponding to the various oscillators. In order to have an half BPS
operator also on the SYM side, these insertions have to be done in all
possible ways. For instance the first entries of the dictionary are
\bea \label{bpsdic}
|0,p^+\rangle & ~\leftrightarrow~~ & \frac{1}{\sqrt{J N^J_o}}
\Ttr[Z^J] = O^J_{\rm vac}~,
\\ \nonumber
\hat{a}_0^{5 \dagger} |0,p^+\rangle & ~ \leftrightarrow~~ & 
\frac{1}{\sqrt{N^{J+1}_o}} {\rm Tr}[\phi^1 Z^J] = O^{J,1}~,
\\ \nonumber
\hat{a}_0^{5 \dagger} \hat{a}_{0}^{6 \dagger}
|0,p^+\rangle & ~ \leftrightarrow~~ & \frac{1}{\sqrt{N_o^{J+2} (J+2)}}
\sum_{l=0}^{J} {\rm Tr}[\phi^1 Z^{l} \phi^2 Z^{(J-l)}]= O^{J,12}
~.
\eea
The normalizations have been put for later convenience: 
% $N_o= \lambda/(8 \pi^2)$. 
$N_o= \Gamma(\omega-1) \frac{\lambda}{8 \pi^\omega}$ and $\omega$ is
related to the space-time dimension $2 \omega=d$.
The relation Eq.\eq{bpsdic} is so natural that one might think that it
was the only possibility. However this is not true, since there are
many other field theory operators with the same quantum numbers, like
$:\Ttr[Z^l] \Ttr[Z^{J-l}]:$ or other possible multi--trace
generalizations. There is a simple reason explaining why the single
trace operator have a special status: as already suggested in the
original paper~\cite{Berenstein:2002jq}
% (see pag. 13, after Eq.~4.10):
the long ``string'' of $Z$'s in the SYM operators {\em is identified
  with the physical IIB string}. The cyclicity of the trace makes the
``string'' of $Z$'s closed; moreover $\a' p^+$ can be also seen as
parametrizing the length of the IIB string as $J$ sets the length of
the field theory operator and we have $p^+ \sim J$. Physically one can
see the SYM operators as a discretized description of IIB strings,
where each $Z$ carries one bit of the total $p^+$ momentum. The BMN
limit $J \to \infty$ just corresponds to the continuum limit of the
discretized string model provided by the ${\cal N}=4$ SYM.
%%%%%%%%%%%%
\begin{figure} %magnification factor .9
\begin{center}
\begin{picture}(0,0)%
\includegraphics{distri.ps}%
\end{picture}%
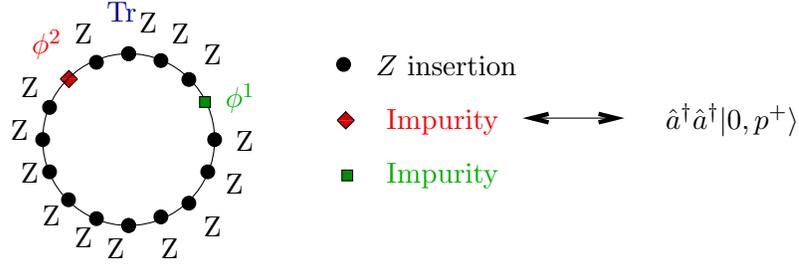
\end{center}
\caption{\label{op-st} The BMN operators provides a discretized model for IIB strings.}
\end{figure}

Up to this point the dictionary is directly inherited from the AdS/CFT
correspondence in the supergravity limit. The new physical input comes
with the identification of the operators corresponding to the excited
string states.  Now, on the string side one can construct the spectrum
by acting on the vacuum with the creation operators
$\hat{a}_n^\dagger,\; \hat b_n^\dagger, \; n\in\zet$. BMN proposed
that the action of these operators corresponds on the field theory
side to the insertion of ``impurities'' in the string of $Z$ as in
Eq.\eq{bpsdic}, but weighted with a $n$--dependent phase. In fact, one
has to insert the appropriate field with a phase proportional to the
level of the string oscillator ($n$) and to the position of the
insertion ($l$) divided by the total number of the fields in the
operator. Finally one has to sum over all possible positions of the
insertion. In this way, one ends up with the following dictionary for
the bosonic states (for $m\not= 0$)\footnote{The phase in Eq.\eq{bosdic} 
is normalized with a factor of $J+2$
simply because this is the number of elementary fields present in the
field theory operator. However, in what follows we will be interested
only in the leading $J\to\infty$ behaviour. See~\cite{Beisert:2002tn} for
an analysis where the phase factors are defined so that the BMN operators
form representations of the superconformal group also at finite $J$.}:
\bea \label{bosdic}
\hat{a}_m^{5 \dagger} |0,p^+\rangle & ~\leftrightarrow~~ & 
\frac{1}{J \sqrt{N_o^{J+1}}} \sum_{l=0}^{J} 
{\rm Tr}[Z^l \phi^1 Z^{J-l}] \ex{2\pi\ii \frac{(l+1) m}{J+2}} = 0 ~,
\\ \label{bosdic2}
\hat{a}_m^{6 \dagger} \hat{a}_{-m}^{5 \dagger}
|0,p^+\rangle & ~\leftrightarrow~~ & \frac{1}{\sqrt{N_o^{J+2} (J+2)}}
\sum_{l=0}^{J} {\rm Tr}[\phi^1 Z^{l} \phi^2 Z^{J-l}]
\ex{2\pi\ii \frac{(l+1) m}{J+2}}~.%,
\eea 
In the following we will refer to this type of $2$-impurity operator
as $O^{J,12}_m$. 
At first sight the Eq.\eq{bosdic} seems strange since it implies that
the state obtained by acting with $\hat{a}_m^{5 \dagger}$ does not
have any corresponding operator on the gauge theory side. But actually
this is correct since also on the string side the state
in Eq.\eq{bosdic} does not belong to the physical spectrum. In fact, as
usual in closed string theory, the physical states have to satisfy the
level matching condition $T \ket{s} = 0 $, where
\beq\label{levelm}
T =  \sum_{n =-\infty}^{+ \infty}
\frac{n}{\a} \left({\hat{a}}^{\dagger}_{n} {\hat{a}}_n + 
{\hat{b}}_{n}^{\dagger} {\hat{b}}_n\right)~.
\eeq
It is interesting to notice that the dictionary
Eqs.\eq{bosdic}-\eq{bosdic2} nicely follows from the pictorial
identification between gauge invariant operators and closed strings
previously discussed.  This interpretation explains also some details
of the correspondence Eq.\eq{bosdic}. In fact one may wonder why on
the left hand (string) side one has to use the hatted oscillators
$\hat{a}_n^{\dagger},\; \hat{b}_n^{\dagger}$ and not the oscillators
$a_n^{\dagger},\; b_n^{\dagger}$ that will be used in the next Section
for the derivation of the $3$--string vertex. The reason is simply
that we used on the field theory side the phase $\exp{[2\pi i
m\frac{l+1}{J+2}]}$ and this phase is exactly the one appearing in the
mode expansion of the bosonic string coordinates in terms of the
hatted oscillators, see Eq.\eq{solut83}! In fact $J$ measures the
length of the SYM operator and so is mapped into $\alpha' p^+ =\alpha$
which measures the length of the string in the light--cone gauge. The
position in the trace labeled with $l$ becomes $\sigma$ ($l \to
\sigma/\a'$) and $m$ is the usual weight for the phase of the $m^{\rm
th}$ string mode. In formulae this implies
\beq
\phi\; \ex{2\pi\ii \frac{m(l+1)}{J+2}} ~\leftrightarrow~
\hat{a}^{\phi \dagger}_m\; \ex{-\ii \frac{m\sigma}{|\a|}} ~~,~~~
\phi\; \ex{-2\pi\ii \frac{m(l+1)}{J+2}} ~\leftrightarrow~
\hat{a}^{\phi \dagger}_{-m}\; \ex{\ii \frac{m\sigma}{|\a|}} ~.
\eeq
Notice that this simple mapping is possible because in the mode
expansion Eq.\eq{solut83} the term proportional to $\ex{-\ii
\frac{m\sigma}{|\a|}}$ involves only $\hat{a}_m^\dagger$
(i.e. $\hat{a}_m^{1 \dagger}$) and not also $\hat{a}_{-m}^\dagger$
(i.e.  $\hat{a}_{m}^{2 \dagger}$). In the fermionic sector the
situation is a bit more complicated. The expansions of the fermionic 
coordinates Eqs.\eq{theta1xx}-\eq{theta2xx} suggest
\bea\label{fermdic}
\psi\; \ex{2\pi\ii \frac{m(l+1)}{J+2}} \,&\leftrightarrow&\, c_m
\left(1 + i \frac{\omega_m-m}{\a \mu}\,\Pi \right )\theta^{\psi
\dagger}_m~, \\ \nonumber % &,&
\psi\; \ex{-2\pi\ii \frac{m (l+1)}{J+2}} \,&\leftrightarrow&\, c_m
\left(1 - i \frac{\omega_m-m}{\a \mu}\,\Pi \right )\theta^{\psi
\dagger}_{-m} ~,
\eea
and similarly for the $\lambda$ impurities. Actually the string
oscillators have here $SO(8)$ spinor indices, while, on the field
theory side, the impurities transform under $SO(4)\times SO(4)$. The
mapping between these two conventions is explicitly spelled out
in~\cite{Pankiewicz:2003kj}.

Let us now focus on the study of the 2--point functions at the planar
level, which already provides non--trivial checks of the dictionary
presented here. On the string side, the states of the Fock space built
with the $\hat{a}^\dagger$'s operators are orthogonal and diagonalize
the free Hamiltonian Eq.\eq{hamil63}. According to Eq.\eq{rela45a},
these masses should translate on the gauge theory side into conformal
dimensions.  However, one should remember that the creation operators
$\hat{a}^\dagger$'s were derived by solving the world--sheet equations
of motion on the {\em whole} complex plane (conformally equivalent to
the sphere). Thus the value of the mass one reads from Eq.\eq{hamil63}
takes into account only the sphere contribution and should be
compared, on the gauge theory side, {\em only} with the contribution
of the planar diagrams. Let us consider an explicit example:
% one derives that {\em at the planar level}
the two--impurity operators Eq.\eq{bosdic2} with different $m$ should be
orthogonal (they are eigenstates of the dilatation operator at this
level) and the corresponding SYM operators should have, {at planar
  level}, the conformal dimensions dictated by Eq.\eq{rela45a}. We
first work in the {\em free} (gauge) theory approximation. It is
easy to see that at this level the conformal dimensions match the
$\mu\a\to\infty$ value of the quadratic Hamiltonian, so let us check
explicitly the orthogonality condition. We can naturally define
a scalar product on the gauge theory side by introducing the following
map:
\beq \label{ob}
\bra{0,p^+} \hat{a}_m^{6} \hat{a}_{-m}^{5} ~\leftrightarrow~~
\lim_{r\to\infty} \bar{O}^{J,12}_n(x) = \lim_{r\to\infty} (r^2)^\Delta
{O}^{* J,12}_n(x)~. 
\eeq
Basically ${O}^{J,12}_n(0)$ and $\bar{O}^{J,12}_n(\infty)$ are
connected by means of the conformal inversion transformation $x'_{\mu}
= x_{\mu} / r^2$, with $\partial x'_{\mu} / \partial x_{\lambda} =
C_{\mu\lambda} (x) / r^2 $, $r\equiv |x|$. This explains also the
factor of $(r^2)^\Delta$ appearing in Eq.\eq{ob}. Thus the scalar product
on the SYM side is
\bea
\lim_{r\to\infty} \langle\bar{O}^{J,12}_n(x){O}_m^{J,12}(0)\rangle &=& 
\frac{1}{(J+2) N^{J+2}_o} 
\sum_{l=0}^J e^{2\pi \ii n\frac{(l+1)}{(J+2)}} 
\sum_{k=0}^{J} e^{2\pi \ii m \frac{(k+1)}{(J+2)}} \nn \\
&\times &\langle \Ttr[\bar{\phi}_1 \bar{Z}^l\bar{\phi}_2\bar{Z}^{J-l}]
\Ttr[\phi_1 Z^k \phi_2 Z^{J-k}]\rangle
\nonumber \\ &=&
\frac{1}{(J+2)} 
\sum_{k,l=0}^J e^{2\pi \ii n\frac{(l+1)}{(J+2)}} 
e^{2\pi \ii m \frac{(k+1)}{(J+2)}}
\delta_{k,J-l} \sim \delta_{n,m}
\label{ortho}
\eea
The second line is obtained by contracting first $\bar{\phi}_1$ and
$\phi_1$ to glue the two traces and then the $\delta$ arises from the
request of contracting  $\bar{\phi}_2$ and $\phi_2$ in a planar way. In
the last step the limit $J\to\infty$ has been taken.
For the vector and fermion impurities the orthonormality condition
is more subtle and requires a suitable definition of the complex
conjugation on the respective operators according to the rules
suggested by the radial quantization \cite{Fubini:1973mf}.
For example, for double--vector
impurities we define~\cite{Chu:2003ji, DiVecchia:2003yp}
\beq
\bar O_{\mu\nu,m}^{J} \equiv \frac{\cN_2}{2}
(r^2)^{\Delta}
\sum_{l=0}^{J}  
\ex{2\pi\ii \frac{(l+1) m}{J+2}} 
\Ttr \left[
(C_{\mu\lambda} \bar{D}_{\lambda} \ r^2\bar{Z} ) \bar{Z}^l 
( C_{\nu\rho}\bar{D}_{\rho}\ r^2\bar{Z} )\bar{Z}^{J-l} 
\right]_x , % (x) \ , \\
\label{baro-ij}
\eeq
where $C_{\mu\lambda}(x) =\delta_{\mu\lambda}-2 x_{\mu}x_{\lambda}/r^2$
is the tensor associated to the conformal inversion transformation. 
The following identities hold for vector
insertions\footnote{$\cN_1=1/{\sqrt{N_o^{J+1}}}$ and
  $\cN_2=1/{\sqrt{N_o^{J+2} (J+2)}}$ are the normalization factors of
  the BMN operators with one and two scalar impurities respectively. These
  have to be divided by one extra $\sqrt{2}$ factor for each vector or fermion
  impurity in order to respect the normalization conditions Eqs.\eq{dzc}-\eq{lac}.}:
\bea\label{dzc}
&&\frac{\cN_1^2}{2}\lim_{r\rightarrow \infty}
\langle0|(r^2)^{\Delta} \Ttr \left[ C_{\mu\lambda}{\del}^\lambda (r^2 \bar{Z})
\bar{Z}^{J} \right](x)  \Ttr \left[ \del'_\nu Z Z^{J} \right](x') |0\rangle = \nonumber \\
&& =  \lim_{r\rightarrow \infty} C_{\mu\lambda} 
\left(\delta_{\lambda\nu} - \frac{2 x_\lambda x_\nu}{r^2}\right) = \delta_{\mu\nu}
\eea
and
\bea
& &\lim_{r\rightarrow \infty}
\langle0| (r^2)^{\Delta} \Ttr
\left[C^{\mu\nu}{\del}_\nu (r^2 \bar{Z}) \bar{Z}^{J} \right](x)
 \Ttr \left[Z^{J} \right] (x')|0\rangle  \nonumber \\
&& =  \lim_{r\rightarrow \infty}
\del_{\mu}\left(\frac{r^2}{(x-x')^2}\right) = 0 \ \ .
\label{dzzc}
\eea
From Eqs.\eq{dzc}--\eq{dzzc} it immediately follows that the
vector impurities in the BMN operators behave exactly as the scalar
impurities, and thanks to the Eq.\eq{ortho} they satisfy the
orthonormality conditions required by the string state/operator
correspondence.

For fermionic impurities we define \cite{DiVecchia:2003yp}
\bea
&&\bar{O}_{\a}^{J} \equiv \frac{\cN_1}{\sqrt{2}} 
\lim_{r\to\infty}
(r^2)^{\Delta}
\Ttr\Big[(\bar{\lambda}\!\not\!\bar x)^\a \bar{Z}^J)\Big](x) \ ,
\label{baro-la} 
\eea  
with
$\not \!\bar x\equiv\bar\sigma_\mu^{\dot\a\a}x^\mu/r$,
$\not \! x\equiv\sigma^\mu_{\a\dot\a}x_\mu/r$.
By using these definitions we have:
\bea
&&\frac{\cN_1^2}{2}\lim_{r\rightarrow \infty}
\langle0|(r^2)^{\Delta} 
\Ttr\Big[(\bar{\lambda}\!\not\!\bar x)^\a \bar{Z}^J)\Big](x)
\Ttr\Big[\lambda_\b Z^J\Big](x')|0\rangle=  \delta^\a_{~\b} \ ,
\label{lac} 
\eea
and similarly for the $\psi$ impurities.
This implies that also the tree-level evaluation
of BMN correlators with fermionic impurities can be reduced 
to the scalar impurity case, apart from some (important) signs due
to the anticommuting nature of $\lambda$ and $\bar\psi$. 
Then, from Eq.\eq{ortho} it follows that also the field theory
operators corresponding to fermionic string excitations are correctly
identified at the tree--level in the planar limit. 
  
Let us now consider the subleading corrections.
To proceed to the one--loop check of the dictionary Eq.\eq{bosdic},
we need the precise form of the couplings in the $\cN=4$ Lagrangian,
which reads in the euclidean space\footnote{We use the euclidean $\sigma$--matrices 
$\sigma^m=({\bf 1},i\tau^i), \bar\sigma^m=({\bf 1}, -i\tau^i)$, where $\tau^i$
are the Pauli matrices, and
$\Ttr(\bar\sigma^m\sigma^n) = 2 \delta^{m n}$.}:
\beqa
L_{E} &=& \frac{1}{g^2_{\rm YM}}\Big(\frac{1}{4}F^a_{\m \n}F^a_{\m \n} + 
(D_\m \bar\phi^I)^a(D_\m\phi^I)^a +
\psi_I^a\sigma^\m(D^\m \bar\psi_I)^a 
+ \lambda^a\sigma^\m(D^\m\bar\lambda)^a
\nonumber\\
&& +\sqrt{2}f^{abc}\left(
\psi_I^a\bar\phi_I^b\lambda^c + \bar\psi_I^a \phi_I^b \bar\lambda^c
\right) 
- \frac{\sqrt{2}}{2}f^{abc}\epsilon_{IJK}\left(
\psi^{I}_a \phi^J_b \psi_c^K + \bar\psi^I_a \bar\phi^{J}_b \bar\psi^{K}_c
\right)
\nonumber\\
&& + f^{abc}f^{ade} \bar\phi_I^b \bar\phi^c_J \phi_I^d \phi_J^e
- \frac{1}{2}f^{abc}f^{ade}\bar\phi^b_K \phi_K^c \bar\phi^d_L \phi_L^e \Big) \ ,
\label{lne}
\eeqa
where $\phi^I\equiv(\phi^1,\phi^2,Z)$ are the three complex scalar
fields of the $\cN=4$ SYM and $\psi^I$ their fermionic superpartners. 
The covariant derivative reads 
$(D_\m \phi)^a = \partial_\m \phi^a + f^{abc}A_\m^b\phi^c$.

\begin{figure} %magnification factor .8
\hspace{.5cm}
\begin{picture}(0,0)%
\includegraphics{gtinte.ps}%
\end{picture}%
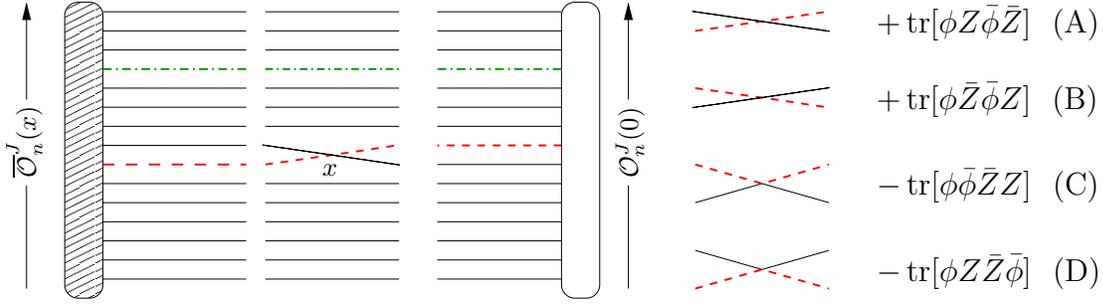
\caption{\label{fig-fterm} The effect of an F-term insertion in a $2$--point function.}
\end{figure}

We consider the 2--point function of BMN operators with two scalar
impurities $\langle \bar O_n^J(x) O_n^J(0) \rangle$. The one--loop
contribution to this correlator is given by the self--energy of the
scalar fields, the gluon exchange and the four scalar interaction
vertices in the last line of Eq.\eq{lne}, which is useful to rewrite
as:
\bea
V_F &=& -\frac{4}{g_{\rm YM}^2}
\Ttr\left(\sum_{i=1}^2 [Z,\phi_i][\bar Z,\bar\phi_i] +
  [\phi_1,\phi_2][\bar\phi_1,\bar\phi_2]\right) 
\label{f-term}\\
V_D &=&  \frac{2}{g_{\rm YM}^2}
\Ttr\left(\frac{1}{2}[Z,\bar Z]^2 + [Z,\bar Z][\phi_i,\bar\phi_i]
+\frac{1}{2} \left(\sum_{j=1}^2 [\phi_j,\bar\phi_j]
 \right)^2\right) \ \ . 
\label{d-term}
\eea
The computation is considerably simplified if we
use the non--renormalization property of the two--point functions of
BPS operators~\cite{D'Hoker:1998tz}.
This property is based on the cancellation between the contribution of
the self--energy, the gluon exchange and the D--term interaction in
Eq.\eq{d-term}.  Since the BPS operators are completely symmetric and
traceless in the scalar fields, the F--term interaction in
Eq.\eq{f-term} does not contribute.  The cancellation between
self--energy, gluon exchange and D--terms can be shown to hold also
for the BMN operators corresponding to true string excitations, as it
is not sensible to the phase factors associated to the impurities and
it is valid term by term in the sum over the position of the
impurities along the string of $Z$
fields~\cite{Kristjansen:2002bb,Constable:2002hw}. The difference with
respect to the BPS (supergravity) operators is in the contribution of
the F--term interaction Eq.\eq{f-term}. In fact, the presence of the
phases associated to the impurities makes these operators no longer
symmetric in the position of the scalar fields.
If we focus on the scalar field $\phi_1$, the effect of the
interaction Eq.\eq{f-term} on the $2$--point functions is summarized by
the Fig.\ref{fig-fterm}. Notice that, as already stressed, we have
to consider only the planar contributions, where the interaction
can connect only contiguous fields.
In particular, the term (A) displaces the impurity $\phi_1$ of one
step to the right, the term (B) moves it one step to the left, while
the remaining two leave the impurity in the same position. The
resulting combinatoric factor will be proportional to the tree--level
result times the respective phases of these four contributions {\it
  i.e.}  $({\rm e}^{2\pi\ii\frac{n}{J}} + {\rm
  e}^{-2\pi\ii\frac{n}{J}} - 2)$.  This picture, where the quantum
corrections have the effect to move the impurities inside the BMN
operators, is consistent with the physical intuition that these
operators are identified with the closed IIB strings. In fact, since
each field on ${O}_n^J$ is seen as a bit of the string, the effect of
the F-term is that of a (discretized) one--dimensional Laplacian
$\partial^2/\partial^2\sigma$. Thus, as 't~Hooft suggested, the (gauge
theory) quantum corrections reconstruct the world-sheet dynamics of
the dual string description and the planarity requirement implies the
locality of the string action.

Concerning now the space--time dependent part, we perform the
computations using dimensional regularization in $2\omega=4-2\epsilon$
dimensions. The configuration space propagator for the scalar fields
is
\be
\langle\bar\phi^{a}_I(x)\phi^b_J(0)\rangle = g_{YM}^2
\delta^{ab}\delta_{IJ} \Delta_{\omega}(x) =  
g_{YM}^2 \delta^{ab}\delta_{IJ}
\frac{\Gamma(\omega-1)}{4\pi^\omega(x^2)^{\omega-1}}
\ \ , 
\label{sc-prop}
\ee
where $\Delta_\omega(x)$ is the Green function for the Laplacian in
$2\omega$ dimensions.
The one--loop contribution to the two--point function between two BMN
operators reads
\bea
\langle {O}^{* J,12}_n(x){O}_n^{J,12}(0)\rangle_{\rm 1loop} &=& 
2g_{\rm YM}^2N \left(\frac{1}{x^{2(\omega-1)}}\right)^{J+2}\!\!\!
I(x)\,({\rm e}^{2\pi\ii\frac{n}{J}} + {\rm e}^{-2\pi\ii\frac{n}{J}} - 2) 
\nn\\
&\sim& - 8\pi^2 n^2 \frac{g_{\rm YM}^2N}{J^2} 
\left(\frac{1}{x^{2(\omega-1)}}\right)^{J+2}\!\!\! I(x)~, 
\label{2pt-1loop}
\eea
where in the last step we have taken the limit $J\to\infty$.
The quantity $I(x)$ in Eq.\eq{2pt-1loop} is the one--loop integral:
\bea
I(x)&=&\left[\frac{\Gamma(\o -1)}{4\pi^\o}\right]^2 (x^2)^{2(\o-1)}
\int d^{2\o}y \frac{1}{(y^2)^{2(\o-1)}[(x-y)^2]^{2(\o-1)}} \nn \\
&=&\frac{1}{8\pi^2}\Big(\frac{1}{\epsilon} + \gamma + 1 + \log\pi
+ \log x^2 + O(\epsilon)\Big) \ \ .
\label{Ix}
\eea
The result in Eq.\eq{2pt-1loop} confirms that the loop expansion
for the BMN operators is actually in terms of the effective parameter
$\lambda'$. Putting together the tree--level and the one--loop result
we get from Eq.\eq{2pt-1loop} and Eq.\eq{Ix}:
\be
\langle{O}^{* J,12}_n(x){O}_n^{J,12}(0)\rangle =
\left(\frac{1}{x^{2}}\right)^{J+2}\!\!\!
%\Delta_\omega(x)^{J+2}  
\left(1-8\pi^2 n^2 \lambda' I(x)\right) \ \ .
\label{tree+one}
\ee
The requirement of the orthonormality among the BMN operators in
planar perturbation theory at one--loop fixes also the finite part in
the renormalization procedure.  In fact, by requiring that the
one--loop renormalized operators 
\be
O_n^{J,12}\equiv
Z(\lambda',\epsilon)O_{n\, {\rm ren}}^{J,12} \ \ ,~~
\label{Oren}
\mbox{with}~~
Z(\lambda',\epsilon)= \left[1 + \frac{n^2\lambda'}{2}
\left(\frac{1}{\epsilon}+ F\right)\right] \ \ ,
\ee
are orthonormalized, we get from Eq.\eq{tree+one}:
\bea
\langle O_{n\, {\rm ren}}^{* J,12}(x)
O_{n\, {\rm ren}}^{J,12}(0)\rangle &=&
\frac{\delta_{nm}}{(x^2)^{J+2+n^2\lambda'}}
\left[1+n^2\lambda'(F-\gamma-1-\log\pi)\right] \nn \\
&\equiv&\frac{\delta_{nm}}{(x^2)^{J+2+n^2\lambda'}} \ \ ,
\label{ortho-one}
\eea
which implies $F=\gamma+1+\log\pi$. Thus the duality with string theory
fixes a precise renormalization scheme for the field theory calculations.

From Eq.\eq{ortho-one} we can read the anomalous
dimension, $\gamma_n= n^2\lambda'$, of the BMN operator $O_n^J$, which
is in agreement with the duality relation in Eq.\eq{rela45a}. This
agreement holds also at higher orders in the quantum expansion, as it
was shown in~\cite{Gross:2002su} by means of an explicit $2$-loop
computation. At first sight this success in reproducing Eq.\eq{rela45a}
with {\em perturbative} gauge theory computations appears strange.
After-all the Penrose limit translates on the SYM side into a strong
coupling limit $\lambda\to \infty$ and, in this regime, the use of
Feynman diagrams is not justified. In principle one should compute
$\langle {O}^{* J,12}_n(x){O}_n^{J,12}(0)\rangle$ exactly in
$\lambda$ and $J$ and only then could one take the double scaling
limit of~Eq.\eq{bmnlimi85}. However, the $2$-loop corrections to
Eq.\eq{2pt-1loop} are proportional to $1/J^4$~\cite{Gross:2002su}
which implies that they do not contribute at first order in
$\lambda'$. This is likely to be a general pattern so that, at the
level of $2$-point functions, the 't~Hooft expansion is automatically
changed term by term into a $\lambda'$-expansion\footnote{However this
  is certainly not the case for correlators with four or more
  operators~\cite{Beisert:2002bb}.}. This expectation is supported by
the analysis of~\cite{Santambrogio:2002sb}, where the the string
prediction Eq.\eq{rela45a} has been beautifully confirmed by a pure field
theory argument.

\section{Further developments}\label{further}

\subsection{The $3$-string vertex}\label{3strsec}

\begin{figure}[htb]
\begin{center}
\begin{picture}(0,0) %magnification factor .75
\includegraphics{3ptvertex.ps}%
\end{picture}%
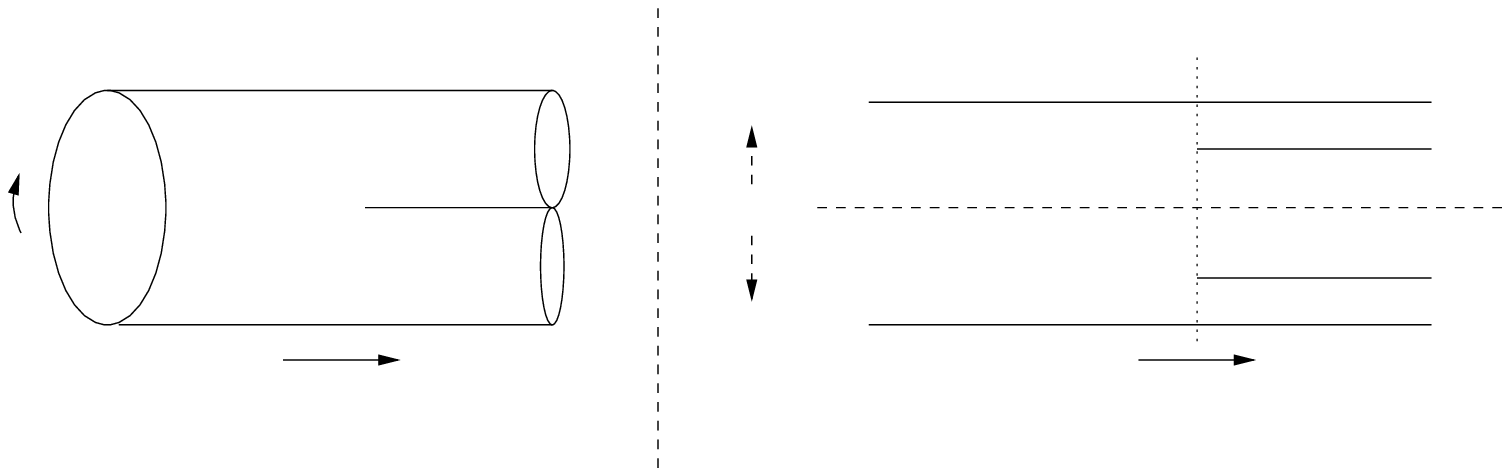
%\vspace{1cm}
%\includegraphics{3ptvertex.eps}
\end{center}
\caption{\label{fig3} The light-cone picture of the three string interaction vertex.}
\end{figure}

As we have seen in Section \ref{freesec}, the quantization of the IIB
superstring on the PP--wave background is straightforward once the
light--cone gauge condition is imposed.  An alternative proposal has
been put forward in~\cite{Berkovits:2002zv}, where the superstring on
the PP--wave background is described in a covariant formalism.  Even
if this looks a promising step toward a covariant quantization, the
presence of a non--trivial R--R field gives rise to a complicated
world-sheet action, and explicit computations of amplitudes in this
framework have not been done yet.  Thus for the computation of the
couplings among the string states we are still obliged to work in the
light--cone. In this gauge, the definition of the usual vertex
operators for string states with $p^+ \ne 0$ is problematic (see
Chapter~7 of~\cite{Green:1987sp}), and a different procedure has to be
applied.  This has been developed
in~\cite{Mandelstam:1973jk,Mandelstam:1974hk,Cremmer:1974jq,Cremmer:1974ej}
for the bosonic string in the flat space and then extended to the
supersymmetric case, always in flat space,
in~\cite{Green:1983tc,Green:1983hw}. This approach consists in
introducing an independent Hilbert space for each 
external state involved in the process, and to describe the
interaction by a state $\ket{H_n}$ in the tensor product of these
Hilbert spaces. The state $\ket{H_3}$ acts as a generating functional
for the $3$-string interactions. In fact, its scalar product with $3$
physical states gives the coupling $C_{3s}$ among the strings under
consideration
\be 
C_{3s}=
(\langle 1 | \otimes \langle 2| \otimes \langle 3|) ~|H_3\rangle \ .
\label{c3}
\ee
Notice that this approach is not particular of the light-cone gauge.
In fact, generating functionals if this type were introduced long time
ago~\cite{Sciuto:1969vz,Caneschi:1969gm} to describe the interactions
in the dual models (covariant bosonic strings, in the modern language).
Moreover this technique is not limited to the computation of the
$3$-point functions, either. At least in principle (and actually in an
explicit way for bosonic strings) it is possible to derive the form of
$\ket{H_n}$.
In addition, each one of these objects has an expansion in topologies,
since the various string interactions are generically corrected at
quantum level. Here we will focus on the simplest case: the $3$-string
interaction at tree-level depicted in Fig.\ref{fig3} and from now on
it is understood that $\ket{H_3}$ indicates only the tree-level
contribution.

In the light-cone gauge, the explicit form of $\ket{H_3}$ is usually
determined by requiring that it satisfies all the symmetries of the
background. This procedure is divided into two steps, reflecting a
natural classification of the symmetries in the light--cone gauge:
first one imposes the {\it kinematical} symmetries, {\it i.e.}  the
symmetries which preserve the light--cone gauge condition, and then
the other {\it dynamical} symmetries.
The outcome of the first step is a ket state $\ket{V}$, 
which for supersymmetric theories has to be completed by a 
{\it prefactor} ${\cal P}$ such that the complete vertex:
\be
\ket{H_3}\equiv {\cal P}\ket{V}
\label{h3}
\ee
satisfies all the kinematical and dynamical symmetries (see Chapter 11
of~\cite{Green:1987mn}). Notice that the light-cone Hamiltonian
itself, being the generator of the $x^+ \to x^+ + c$ transformations,
is a dynamical symmetry, since the $x^+$ shifts clearly do not
preserve the gauge condition $x^+ = e(\a) \tau$. From Fig.\ref{fig3}
it is easy to see a key property of the dynamical generators: they can
not be expressed just in terms of the free dynamics, but get
corrections also from the interaction. In fact a shift from $x^+ =
-\epsilon$ to $x^+ = \epsilon$ makes us pass from a $1$-string state
to a $2$-string state. Because of this in the light-cone gauge the
$3$-point interaction can be seen as non-linear correction of the 
free Hamiltonian and for this reason we have indicated it with
$\ket{H_3}$.

In the case of the PP--wave background, the presence of the discrete
symmetry $\zet_2$, which exchanges the two $SO(4)$'s acting on the
transverse coordinates to the light--cone plane (see Section
\ref{maximally}), makes the derivation of $\ket{H_3}$ more
subtle~\cite{Chu:2002eu}. In fact, in order to have a
$\zet_2$--invariant vertex $\ket{H_3}$ there are two possible choices
for the kinematical vertex $\ket{V}$ and the prefactor ${\cal P}$:
they can be both {\it odd} or both {\it even} under the $\zet_2$
transformation.
Both the possibilities have been considered in the literature.
The first has been studied  
in~\cite{Spradlin:2002ar,Spradlin:2002rv,Pankiewicz:2002gs,
Pankiewicz:2002tg} 
and then in~\cite{Pankiewicz:2003kj,Pankiewicz:2003ap}, where an
equivalent formulation of this vertex was given and its transformation
properties under the $SO(4)\times SO(4)\times \zet_2$ symmetry group
were clarified. The second possibility was analyzed instead
in~\cite{Chu:2002eu,Chu:2002wj,DiVecchia:2003yp}.

Here we sketch a derivation of the interaction vertex based on
standard path integral techniques\footnote{This approach has been
  worked out in collaboration with P. Di Vecchia, J.L. Petersen and
  M. Petrini.}. Since the tree-level string dynamics is captured by a
set of harmonic oscillators, we can use the usual transition
amplitude~\cite{Faddeev:1975vr,IZ} in the coherent state phase space
with variables $(\ra,\ra^*)$\footnote{Notice that $(\ra,\ra^*)$
are in this formalism complex numbers corresponding to the eigenvalues
of the string oscillators $(a,a^\dagger)$.
$\ra^*$ is an arbitrary complex number and not necessarily 
the complex conjugate of $\ra$.
The same holds for the fermionic phase space variables $(\rb,\rb^*)$
that we will shortly introduce.}  
\begin{eqnarray}\nn
&&U(\ra^*,\ra,t''-t')=\int \prod_l \frac{d\ra^*(t_l)d\ra(t_l)}{2\pi i}
\exp\{\frac{1}{2}(\ra^*(t'')\ra(t'')+\ra^*(t')\ra(t'))\} \\
&&\times\exp\left \{ i\int_{t',\ra}^{t'',\ra^*}
\left ( \frac{1}{2i}(\dot{\ra}^*(t)\ra(t)-
\ra^*(t)\dot{\ra}(t) - H(\ra^*(t),\ra(t))\right ) dt\right \}~,
\label{oscillator-p-i}
\end{eqnarray}
where $l$ enumerates the intermediate times resulting in the path integral
when their number grows to infinity.
The limits of the path integral indicate that the variable $\ra(t)$ is
fixed at time $t'$ and the variable $\ra^*(t)$ is fixed at time $t''$.
$H(\ra^*,\ra)$ is the classical
Hamiltonian Eq.\eq{hamil63}.
The object
integrated over $t$ is seen to be the phase space action, which 
vanishes here due to the equations of motion.
The final form of the transition amplitude
is then
\be
\exp\left \{\frac{1}{2}\left (\sum_{\mbox{\footnotesize initial oscillators}}
(\ra_m^*\ra_m)
+ \sum_{\mbox{\footnotesize final oscillators}}(\ra_m^*\ra_m)
\right )\right \}~.
\label{trans}
\ee
One easily verifies that the matrix element of the
evolution operator takes the same form for a fermionic oscillator with
phase space variables $(\rb,\rb^*)$.

The dynamics is governed by the classical equations of motion except
for the interaction point that we choose to be $x^+=0$.  The
fields on the string world sheet are described by different sets of
free oscillators for $x^+ >0$ and $x^+ <0$. In the
parametrization chosen in Fig.\ref{fig3} the evolution for $x^+ <0$ is
determined only in terms of the bosonic and fermionic modes
$\ra^{(3)}_n, \ra^{*(3)}_n$, $\rb^{(3)}_m, \rb^{*(3)}_m$ of the third
string, while for $x^+ >0$ is written in terms of the modes
$\ra^{(i)}_n, \ra^{*(i)}_n$ and $\rb^{(i)}_m, \rb^{*(i)}_m$, of the
other two strings $i=1,2$.  The interesting part of the problem
concerns the transition from one string at $x^+ = -\epsilon$ to two
strings at $x^+ = +\epsilon$ with $\epsilon$ a small positive time.
During that transition we demand the world sheet to be continuous and
smooth, {\em i.e.} we demand, as usual, the continuity of the phase
space trajectories.
This condition yields some relations among the modes in the form of a
linear system.  Solving this linear system by choosing a set of
independent variables and substituting back in Eq.\eq{trans} one gets
the final form of the vertex.  Notice that by applying this procedure
one gets only the exponential part $\ket{V}$ of the vertex Eq.\eq{h3}.
In the path integral formalism the prefactor ${\cal P}$ should come
from extra contributions related to the gauge fixing of the
$\kappa$--symmetry at the point on the world sheet where the 3 strings
join. By ignoring this complication, we get exactly the kinematical
part of the vertex. 
To see this, let us consider first the bosonic modes: the linear
system is obtained by rewriting the continuity conditions:
\beq
x_{m(3)} = - \sum_{n = -\infty}^{\infty} \sum_{\ri=1}^{2}
\frac{\alpha_\ri}{\alpha_3 }X^{(\ri)}_{mn} 
x_{n(\ri)} ~~,~~
p_{m(3)}  =  -  
\sum_{n = -\infty}^{\infty} \sum_{\ri=1}^{2} X^{(\ri)}_{mn} p_{n(\ri)} 
\label{eq378}
\eeq
by means of
Eqs.\eq{xxx53} and \eq{ppp73}. 
Then, by choosing the $\ra^*$ modes as
independent variables, one can solve for the $\ra$ modes getting:
\begin{equation}
\ra^{(r)}_n = \sum_{s=1}^{3} \sum_{m
=-\infty}^{\infty} N^{rs}_{nm}\ra^{(s)*}_m~.
\label{arxxx}
\end{equation}
From Eqs.\eq{arxxx} and \eq{trans} we then get: 
\be
\exp{\left[\frac{1}{2}\sum_{r,s=1}^{3} \sum_{n,m = -\infty}^{\infty}
\ra_{n}^{(r)*} N^{rs}_{nm} \ra_{m}^{(s)*}\right]}=
{}_{123}\bra{v}{\rm e}^{\ra_n^{(r)*}a_n^{(r)}}\ket{H_3}~.
\label{ver45}
\end{equation}
For the definition of the $X$ matrices in Eq.\eq{eq378} and the explicit form of
the Neumann matrices $N^{rs}_{nm}$ we refer to~\cite{Pankiewicz:2002gs}. 
The state $\ket{V}_{b}$ in the Hilbert space corresponding
to the wave--function in Eq.\eq{ver45} is exactly the 
bosonic part of the vertex worked out in \cite{Spradlin:2002ar} 
by imposing the kinematical constraints
as Dirac $\delta$--functions in the momentum space.

Concerning now the fermions, the ambiguity 
due to the presence of the $\zet_2$ symmetry 
that we mentioned before shows up in the path
integral formalism as the possibility to choose {\it different boundary
conditions} in the zero mode sector. 
In~\cite{Spradlin:2002ar,Spradlin:2002rv,Pankiewicz:2002gs,
Pankiewicz:2002tg}, closely following the flat space analysis
of~\cite{Green:1983hw}, the fermionic zero modes are treated on
different footing with respect to the others modes, as they
are represented in terms of the null eigenstate $\ket{0}$ of the fermionic 
coordinate zero--mode $\theta_0$:
\be
\theta_0\ket{0}=0~.
\label{fvacuum}
\ee
Thus the boundary conditions in the zero--mode sector are specified
in terms of the $(\theta_0,\lambda_0)$ phase space variables, instead
of the coherent state variables $(\rb_0,\rb_0^*)$.
By applying the same procedure discussed before for the bosonic case
and choosing $\lambda_0$ as the independent variables for
the zero--mode sector, one ends up with the state:
\bea
\ket{V}_f^I&=&\exp\left[\sum_{r,s=1}^3
\sum_{m,n=1}^{\infty}b^{\dagger}_{-m(r)}{\cal Q}^{rs}_{mn}b^{\dagger}_{n(s)}
-\sqrt{2}\L\sum_{r=1}^3\sum_{m=1}^{\infty}{\cal Q}^r_m 
b^{\dagger}_{-m(r)}\right]
\delta_0\ket{0}_{123}~, \nn\\
\label{vf1}
\eea
where $\delta_0=\prod_{a=1}^8(\sum_{\rr=1}^3\lambda_0^{a(\rr)})$
is the delta--function imposing the conservation of the zero--mode 
fermionic momentum.
The ${\cal Q}$--matrices appearing in Eq.\eq{vf1} are given by:
\bea
\label{qmn}
{\cal Q}^{rs}_{mn} & =&e(\a_r)\sqrt{\left|\frac{\a_s}{\a_r}\right|}
\bigl[P_{(r)}^{-1}U_{(r)}C^{1/2}N^{rs}C^{-1/2}U_{(s)}
P_{(s)}^{-1}\bigr]_{mn}\,,\\
\label{qm}
{\cal Q}^r_n & =&\frac{e(\a_r)}{\sqrt{|\a_r|}}(1-4\m\a
K)^{-1}(1-2\m\a K(1+\Pi))\bigl[P_{(r)}C_{(r)}^{1/2}C^{1/2}
N^r\bigr]_n\,,\\
\L & =& \a_1\l_{0(2)}-\a_2\l_{0(1)} \quad , \quad \a \equiv \a_1\a_2\a_3~.
\eea
Eq.\eq{vf1} is exactly the kinematical part of the fermionic vertex 
written in~\cite{Pankiewicz:2002gs}, to which we refer
for the definition of the quantities appearing in Eqs.\eq{qmn}-\eq{qm}.
Notice that the state $\ket{0}$ in Eq.\eq{fvacuum} is not the vacuum state
for the string theory on the PP--wave, as
$H_{(\rr)}\ket{0}=4\mu\alpha_{\rr}\ket{0}$.
Nonetheless, it reduces to the flat space vacuum in the $\mu\to 0$ limit.
Thus this construction ensures a smooth limit of the interaction vertex 
to the flat space.

An alternative proposal has been made in~\cite{Chu:2002eu,
Chu:2002wj,Chu:2003nb}, where the vertex is built on the
true vacuum of the theory $\ket{v}$, $b_0\ket{v}=0$.
In this case, the fermionic modes $b_0, b_0^{\dagger}$ are treated on the same
footing as the other modes, since, contrary to the flat space case,
they have a non--zero energy and they are not true zero--modes of the 
Hamiltonian Eq.\eq{hamibb}. Thus the coherent state variables 
$(\rb_0,\rb_0^{*})$ are used for the zero--modes as well as for all 
the other modes $m \ne 0$,
and the resulting state is~\cite{Chu:2002wj,Chu:2003nb}:
\bea\label{E-la} 
\ket{V}_f^{II} & = & \exp
\left\{ \frac{1 + \Pi}{2}
\left[\sum_{\rr,\rs=1}^3\sum_{m,n=1}^{\infty} b^{\dagger}_{-m(\rr)}
Q_{mn}^{\rr\rs} b^{\dagger}_{n(\rs)} - \sqrt{\a'} \Lambda
\sum_{\rr=1}^3\sum_{m=1}^{\infty} Q^\rr_m b^{\dagger}_{-m(\rr)}
\right] \right.\nonumber \nonumber \\ & & \quad+~\left. \frac{1 -
\Pi}{2} \left[\sum_{\rr,\rs=1}^3 \sum_{m,n=1}^{\infty}
b^{\dagger}_{m(\rr)} Q_{mn}^{\rr\rs} b^{\dagger}_{-n(\rs)} +
\frac{\alpha}{\sqrt{\a'}} \Theta \sum_{\rr=1}^3\sum_{m=1}^{\infty}
Q^\rr_m b^{\dagger}_{m(\rr)} \right] \nonumber \right\} \\ & &
\times~\exp \left\{- \sum_{\ri=1}^2
\sqrt{\frac{\alpha_{\ri}}{|\alpha_3|}} b^{\dagger}_{0(\ri)}
b^{\dagger}_{0(3)} \right\} \ket{v}_{123}~, 
\eea 
where 
\be 
\Theta \equiv \frac{1}{\a_3} (\th_{0(1)}- \th_{0(2)})~.
\ee 
The matrices $Q$ are diagonal in the spinor space and are defined as:
\bea 
&&Q^{\rr\rs}_{mn}\equiv e(\a_\rr) \sqrt{\frac{|\a_\rs|}{|\a_\rr|}}\;
    [U_{(\rr)}^{1/2} C^{1/2} N^{\rr\rs} 
C^{-1/2} U_{(\rs)}^{1/2}]_{mn},\label{defQ1} 
\\ && \label{defQ2}~
Q^{\rr}_m \equiv \frac{e(\a_\rr)}{\sqrt{|\a_\rr|}} [U_{(\rr)}^{1/2}
  C_{(\rr)}^{1/2} C^{1/2} N^\rr]_m~.  
\eea 
Notice that using the relation 
$P^{\pm 2}_{n (r)} = U_{n (r)} + \frac{\a_r \mu}{n} (1\mp \Pi)$
one can show that the matrices ${\cal Q}$ and $Q$ appearing
respectively in Eq.\eq{vf1} and Eq.\eq{E-la}
coincide for the non--zero modes with positive chirality
$b_n^{a\dagger}$, $n\ne 0$, $a=1,\ldots,4$. 

Let us now come to the supersymmetric completion of the kinematical
vertices discussed so far. The {\it dynamical}
supersymmetry charges of the PP--wave background are given by
$Q^-,\bar Q^-$ from which we define the combinations
\begin{equation}
Q = \frac{1}{\sqrt{2}} \left(Q^- + \bar{Q}^- \right)~~,~~
\widetilde{Q} = \frac{i}{\sqrt{2}} \left( Q^- - \bar{Q}^- \right)~.
\end{equation}
In the $\mu\to 0$ limit $Q$ contains just left
moving oscillators, while $\widetilde{Q}$ depends only on the right
moving ones so that $Q$ and $\widetilde{Q}$ are the direct
generalization of the supercharges usually considered in flat--space
computations. 
These charges satisfy the algebra
\begin{eqnarray}\label{sal1}
& \{Q_{\dot{a}},Q_{\dot{b}}\}=
2 \delta_{\dot{a} \dot{b} } (H+T) ~~,~~~
\{\widetilde{Q}_{\dot{a}},\widetilde{Q}_{\dot{b}}\}=
2 \delta_{\dot{a} \dot{b} } (H-T)&
\\ \label{sal2} %\nonumber
& \{Q_{\dot{a}},\widetilde{Q}_{\dot{b}}\}=
\mu\left[- (\gamma_{ij}\Pi)_{\dot{a}\dot{b}}J^{ij}+
(\gamma_{i'j'}\Pi )_{\dot{a}\dot{b}}J^{i'j'}\right]~,&
\end{eqnarray}
where $T$ is the operator defined in Eq.\eq{levelm}, which is vanishing
on the physical states. 
As already remarked,
in order for the full supersymmetry algebra to be satisfied at the
interacting level, the kinematical vertex has to be completed with
a polynomial prefactor. An analogous construction also applies
for the dynamical supersymmetry generators $Q$ and $\widetilde{Q}$.
To our knowledge there is no way to
derive the prefactor from first principles, the standard approach
being to write a suitable ansatz and then check that it is invariant
under all symmetries~\cite{Green:1983tc,Green:1983hw}.
To proceed further, some physical inputs are then required.

In~\cite{Spradlin:2002ar,Spradlin:2002rv,Pankiewicz:2002tg} 
the continuity of the interaction vertex
in the $\mu\to 0$ limit is required. 
The vertex is thus built starting from the kinematical part
\be
\ket{V}^I \equiv \ket{V}_b \otimes \ket{V}_f^I
\delta\left(\sum_{\rr=1}^3 \a_\rr\right)
\label{kv1}
\ee
given by the tensor product of the states in Eq.\eq{ver45} and Eq.\eq{vf1}.
The prefactor turns out in this
case to be a slight modification of the flat space prefactor
computed in~\cite{Green:1983hw}, and the full
Hamiltonian and the dynamical charges read~\cite{Pankiewicz:2002tg}:
\bea
\label{h}
\ket{H_3}^I & = &
\left((1-4\m\a K)\widetilde{\cal K}^I{\cal K}^J-\m\frac{\a}{\a'}\d^{IJ}\right)
v_{IJ}(Y)\ket{V}^I\,,\\
\label{q}
\ket{Q_{3\,\dot{a}}}^I & = & (1-4\m\a K)^{1/2}\widetilde{\cal K}^I
s_{\dot{a}}^I(Y)\ket{V}^I\,,\\
\label{tq}
\ket{\widetilde{Q}_{3\,\dot{a}}}^I & = & (1-4\m\a K)^{1/2}{\cal K}^I
\tilde{s}_{\dot{a}}^I(Y)\ket{V}^I\,.
\eea
Here ${\cal K}, \tilde{\cal K}$ and $Y$ are respectively the bosonic
and fermionic constituents of the prefactor. They are found by
requiring the commutation with the kinematical constraints, such that
the states in Eqs.\eq{h},\eq{q},\eq{tq} still satisfy them. Again, in
the $\mu\to 0$ limit ${\cal K}$ ($\tilde{\cal K}$) depends only on the
left (right) moving oscillators.
The explicit expression of these prefactor constituents is reported 
in~\cite{Spradlin:2002ar,Pankiewicz:2002tg}, where one can find also
the explicit form of the functions $v_{IJ}, s_{\dot a}^I, \tilde s_{\dot a}^I$ 
as an expansion in powers of the Grassmann variables $Y$.
In~\cite{Pankiewicz:2003kj,Pankiewicz:2003ap}, it was shown that a
vertex completely equivalent to that in Eq.\eq{h} can be obtained by
starting from the kinematical vertex Eq.\eq{E-la}, and, of course, by
insisting on the continuity of $\mu\to 0$ limit. This requirement
forces us to assign an even $\zet_2$ parity to the state
$\ket{0}$~\eq{fvacuum}. In this case the vacuum has to be
$\zet_2$--odd because $\ket{v}$ and $\ket{0}$ have opposite
parity~\cite{Chu:2002eu}. The $\mu\to 0$ smoothness is the main
argument for this parity assignment since it is also used in the
supergravity analysis of~\cite{Kiem:2002pb}.

A different approach was suggested
in~\cite{Chu:2002eu,Chu:2002wj,Chu:2003nb}: following the gauge theory
intuition, the vacuum state of the string Fock space is defined to be
{\it even} under the discrete $\zet_2$ symmetry even if this is in
contrast to what the continuity of the $\mu\to 0$ limit requires. This
assignment is not only natural because of the form of $O^J_{\rm
  vac}$~\eq{bpsdic}, but it also necessary on the SYM side. In fact
the overlap between three vacuum operators is not vanishing ({\em
  i.e.} of the same order as all the others correlators), and this is
not consistent with the selection rule implied by the odd-parity
previously considered. Thus we are led to give up the continuity of
the flat space limit $\mu\to 0$ for the string interaction. Also in
this case it is possible to build a string vertex that is invariant
under the $\zet_2$ transformation, thus realizing this symmetry
explicitly ({\em i.e.} both the interaction and the vacuum state are
$\zet_2$ invariant at the same time). As we will see in the next
section, this approach is consistent with the proposal
of~\cite{Constable:2002hw,Dobashi:2002ar} for the comparison with the
field theory.  In this case, one starts from the kinematical vertex
\be
\ket{V}^{II}\equiv\ket{V}_b\otimes\ket{V}_f^{II}
\delta\left(\sum_{\rr=1}^3 \a_\rr\right)
\label{kv2}
\ee
given by the tensor product of the bosonic state in Eq.\eq{ver45} and
the fermionic one given in Eq.\eq{E-la}. The supersymmetric completion
for this vertex turns out to be very simple~\cite{DiVecchia:2003yp}:
\bea\nn 
&&\ket{H_3}^{II} = \sum_{\rr=1}^3 H_\rr \ket{V}^{II}~~,~~ 
\ket{Q_{3\dot{a}}}^{II} = \sum_{\rr=1}^3 Q_{r \dot{a}}\; \ket{V}^{II}~~,~~
\ket{\widetilde{Q}_{3\dot{a}}}^{II} = \sum_{\rr=1}^3 \widetilde{Q}_{\rr
\dot{a}}\; \ket{V}^{II}~.\\
\label{nqh}
\eea
With this ansatz the relations Eqs.\eq{sal1}--\eq{sal2} hold also at
the interacting level as a direct consequence of the free--theory
algebra. This remarkably simple form has a very similar structure to
what was proposed in the string bit formalism~\cite{Verlinde:2002ig}.
Moreover the ansatz Eq.\eq{nqh} shares some striking similarity also
with the behaviour of supergravity on $AdS_5 \times S^5$. The vertex
derived here is related to the cubic {\em bulk} couplings derived from
the compactification of IIB theory on $AdS_5\times
S^5$~\cite{Kiem:2002xn}. Since the PP--wave background is obtained as
a Penrose limit from the $AdS_5\times S^5$ geometry, we should be able
to compare the results of $\ket{H_3}$ for supergravity states with the
(leading order in $J$) results obtained in $AdS_5\times S^5$. It is
interesting to notice that the bulk vertices obtained
in~\cite{Lee:1998bx} for $3$ scalars are indeed proportional to
$\Delta_1 + \Delta_2 - \Delta_3$, exactly as the $3$--point functions
derived from $\ket{H_3}^{II}$.

Let us summarize the results reported in this subsection.
We have seen that, due to the lack of a covariant
quantization of the string on the PP--wave background
Eq.\eq{pwave-back}, the determination of a supersymmetric
$3$-string interaction vertex requires some extra physical inputs.
This lead to an ambiguity in the construction of the vertex,
and two inequivalent solutions have been proposed.
%\noindent
The first is obtained by requiring the continuity of the
flat space limit $\mu\to 0$ and results in the vertex
given by Eq.\eq{h}~\cite{Spradlin:2002ar,Spradlin:2002rv,Pankiewicz:2002gs,
Pankiewicz:2002tg}. The same approach is followed in~\cite{Pankiewicz:2003kj}
which discusses the $SO(4)\times SO(4)$ formulation 
of the vertex.
These two formulations of a string vertex smoothly connected
to the flat space have been shown to be equivalent 
in~\cite{Pankiewicz:2003ap}.  
%\noindent
The second solution more closely follows the behaviour of supergravity
in $AdS_5\times S^5$. The idea is that, since the PP--wave background
can be seen as an approximate description of $AdS_5 \times S^5$, even
for small curvatures, the 3--state interaction has to be compared with
the results in $AdS_5 \times S^5$ rather than with those of
flat--space.

\subsection{The operator mixing}\label{mixsect}

In the previous section we introduced the BMN operators and showed
with an explicit $1$-loop computation that they have, {\em at the
  planar level}, a well defined conformal dimension. However the
conformal dimensions will in general receive contributions also from
non-planar diagrams. Again, at $1$-loop, one can focus just on the
F-term interaction Eq.\eq{f-term}, since the contributions of all
other vertices cancel among
themselves~\cite{Kristjansen:2002bb,Constable:2002hw}. The main
difference with the diagrams previously considered is that now we have
to take into account interactions between non-contiguous fields.
\begin{figure}[tp]
\begin{center}
\begin{picture}(0,0)%
\includegraphics{mixing.ps}%
\end{picture}%
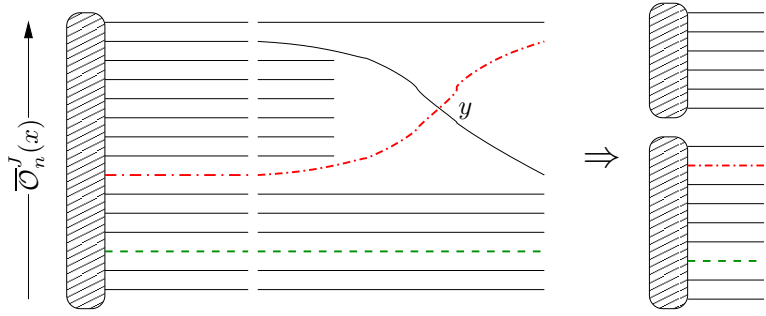
\end{center}
\caption{\label{nonpl-fig} As effect of ``non-local'' interaction, a single trace
  operator splits in two. }
\end{figure}
% \noindent
Thus, while the space-time part of the computation is basically the
same as before, in this case the contraction between the two
non-contiguous fields splits the BMN operator ${O}^J_{m}$ in two
traces, one of length $l$ and the other of length $J-l$, see
Fig.\ref{nonpl-fig}.  From this picture, it is quite natural to
expect that non-planar quantum corrections will mix single and double
trace operators and that the true states with definite conformal
dimension will be some linear combination of the two. At the
conceptual level, the simplest way to check this is to generalize the
perturbative computation done in Section~\ref{bmndsect}. We need to
consider the single trace operators (like Eqs.\eq{bpsdic} and \eq{bosdic2})
together with the double-trace states that can be built by multiplying
the BMN operators among themselves (for instance $:O^{J'}_{k}
\Ttr[Z^{J-J'}]:$). Let ${\cal O}_\a$ indicate a generic operator in
this set with two impurities, then the $2$-point function takes the
form
\beq\label{ST}
\langle{\cal O}_\a^{* J,12}(x) {\cal O}_\beta^{J,12} (0) \rangle = 
\left(\frac{1}{x^2} \right)^{J+2} \left( S_{\a\beta} +
T_{\a\beta} \log|x \Lambda|^{-2}\right)
\eeq
The matrices $S$ and $T$ can be computed perturbatively, as was first
done in~\cite{Kristjansen:2002bb,Constable:2002hw}, where the overlaps
among the original BMN operators were considered. However in order to
derive the eigenstates of the dilatation operator at the non-planar level, 
we need also the
overlaps between single and multi-trace
operators~\cite{Beisert:2002bb,Constable:2002vq}. At this point it is
sufficient to diagonalize $T$ in the inner product given by $S$ and, in
particular, to extract the eigenvalues of $T$. These are the conformal
dimensions including the first non-planar correction. 
Let us briefly describe the ideas
of~\cite{Beisert:2002ff} that provide a much simpler
and more intuitive way to derive the same results. In the planar case,
the quantum correction depicted in Fig.\ref{fig-fterm} was
interpreted as a discretized Laplacian, whose action on the BMN
operators was basically that of multiplying the original operator by
$\lambda ({\rm e}^{2\pi\ii\frac{n}{J}} + {\rm e}^{-2\pi\ii\frac{n}{J}}
- 2)$.  In a similar manner, we can interpret the non-planar
contractions as an operation in the set of operators ${\cal O}_a$ that
acts non-diagonally on the single and double-trace states. In fact, we
can quantitatively summarize the process of Fig.\ref{nonpl-fig} by
means of the ``operator'' $h_+$~\cite{Beisert:2002ff}
\beq\label{h+}
h_+ O^{J,12}_{m} = g_2 \lambda' \sum_{k=-\infty}^\infty \sum_{J'=0}^{J} 
\frac{k\sqrt{1-r}\sin^2(\pi m r)}{\sqrt{J} \sqrt{r}\pi^2 (k-m r)}
O^{J',12}_{k} O^{J-J'}_{\rm vac}~,
\eeq
where $0<r=J'/J<1$. This can be derived by rewriting the phase
$\exp{[2\pi i m (l+1)/(J+2)]}$ of the original BMN operator
$O^{J,12}_{m}$ in terms of the phase factors containing $J'$ which are
appropriate for the shorter operator $O^{J',12}_{k}$. Analogously also
the opposite phenomenon is possible and two traces can be glued into a
single-trace operator, which is usually indicated with
$h_-$~\cite{Beisert:2002ff}.  Now the basic idea is to see the action
of the quantum corrections on the generalized BMN operators ${\cal
  O}_\a$ as an Hamiltonian in a quantum mechanical system. In
particular we can treat $h_+$ and $h_-$ as perturbations with 
respect to the planar contribution, since they are subleading in the
parameter $g_2$. By applying the usual formulae of non-degenerate
perturbation theory , one can easily see that there is no ``energy''
shift ({\em i.e} no modification of the conformal dimensions) at the
first order in $g_2$, since $(h_+ + h_-)_{mm}=0$.  However, from the
usual first order formula $\sum_{n\neq m} (h_+ + h_-)_{mn}/(E_m-E_n )|n\rangle$, we get a
non-trivial redefinition of the eigenstates %. For instance
\beq\label{chp}
:O^{\prime\, J,12}_{m}: = 
:O^{J,12}_{m} :
-\sum_{J',k}\frac{g_2\, r^{3/2}\sqrt{1-r}\,\sin^2(\pi m r)\, 
{k}}{\sqrt{J}\,\pi^2(k-m r)^2 {(k+mr)}}\,
:O^{J',12}_{k} O^{J-J'}_{\rm vac} :~.
\eeq
Notice that the factor of $\lambda'$ present in Eq.\eq{h+} and typical
of any $1$-loop computation is not present in Eq.\eq{chp} because
$(E_m-E_n ) \sim \lambda' (m^2-n^2/r^2)$. This approach can be pushed
to the next orders and the conformal dimensions at order $g_2^2$ can
be derived~\cite{Beisert:2002ff} from the usual formula, $\delta
\Delta_n^{(2)}= \sum\limits_{m\not= n} |\langle n|(h_+ + h_-)
|m \rangle|^2/ (\Delta_n^{(0)} - \Delta_m^{(0)})$, obtaining
\beq\label{dg2}
\Delta_m = J+2+\lambda'\left\{
m^2 +\frac{g_2^2}{4\pi^2}\left( \frac{1}{12} +\frac{35}{32\pi^2m^2}
\right)\right\}~.
\eeq
Two remarks are in order here. First, to obtain Eq.\eq{dg2} we did not
sum over all the contribution $m$, but we restricted ourselves to the
two-impurity operators, neglecting the impurity
non--preserving amplitudes. Then a word of caution is needed on the
applicability of the non-degenerate perturbation theory to the second
order. Even if no problems seem to be present in the computation that
leads to the energy shift Eq.\eq{dg2}, the formula that should give
the $g_2^2$-corrected eigenstates breaks down. This is due to the
existence at this order of a non-trivial overlap between single and
triple-trace operators which can not be disregarded (as it has been
done here in the derivation of Eq.\eq{dg2}). This issue has been
studied in~\cite{Freedman:2003bh}, where it is also suggested that the
breakdown of non-degenerate perturbation theory implies that the
string states on the PP-wave are generically unstable.

\subsection{The string/gauge theory dictionary revisited}\label{bmnrev}

The main question now posed by the analysis of the previous section is
whether (and how) the operator mixing is relevant for the BMN
dictionary between string states and gauge invariant operators. This
is particularly interesting because the very same pattern is present
also in the full AdS/CFT correspondence, where the role of the
parameters $\lambda'$ and $g_2$ is played by $\lambda$ and $1/N$.
Conceptually the most simple solution would be to relate the string
states and the true eigenstates of the dilatation operator so that the
r.h.s.~of Eq.\eq{bosdic2} is substituted by $O_m^{' J}$ of
Eq.\eq{chp}.  A lot of work has been done in this
direction~\cite{Janik:2002bd,Chu:2003qd,Georgiou:2003aa,Chu:2003ji},
but eventually it turned out that neither of the approaches to the
string interaction discussed in Section~\ref{3strsec} could be easily
used together with this proposed string/gauge theory mapping. Because
of this, a less stringent requirement on the dictionary was first
proposed in the context of the string bit model~\cite{Verlinde:2002ig}
and then applied in the field theory calculations
of~\cite{Gross:2002mh,Gomis:2002wi}: in order to find the SYM
operators corresponding to the string states, it is necessary to
diagonalize only $S_{\a\beta}$ in Eq.\eq{ST}, because this represents,
on the gauge theory side, the scalar product among the states in the
string Fock space. Of course, this diagonalization can be done in many
different ways, since two bases related by an orthogonality rotation
have the same scalar product; for example, the basis of the true
conformal operators satisfies this requirement, since in this case
both $S_{\a\beta}$ and $T_{\a\beta}$ are diagonal.
A different choice was proposed in the string bit
model~\cite{Vaman:2002ka,Pearson:2002zs} and discussed
in~\cite{Gross:2002mh,Gomis:2002wi} from the gauge theory prespective.
It was proposed that, at the first order in $g_2$, the operators dual
to the string states are
\bea
\Omega^{J,12}_{m} ~=~ 
O^{J,12}_{m}
& - & \sum_{J',k}\frac{g_2\, r^{3/2}\sqrt{1-r}\,\sin^2(\pi m r)}
{{2} \sqrt{J}\,\pi^2(k-m r)^2}
\, O^{J',12}_{k} O^{J-J'}_{\rm vac}
\nonumber \\ \label{cob} & + & 
\sum_{J'}\frac{g_2\,\sin^2(\pi m r)}
{{2} \sqrt{J}\,\pi^2 m^2}  O^{J',1} O^{J-J',2}
\\ \nonumber \Omega^{J',12}_{k} \Omega^{J-J'}_{\rm vac} 
& \!\!\!\!\!\!\!  = & \!\!\!\!\!\!
O^{J',12}_{k} O^{J-J'}_{\rm vac} - 
\sum_{m}\frac{g_2\, r^{3/2}\sqrt{1-r}\,\sin^2(\pi m r)}
{{2} \sqrt{J}\,\pi^2(k-m r)^2}
\, O^{J,12}_{m}
\\ \nonumber \Omega^{J',1} \Omega^{J-J',2} 
%& = & 
& \!\!\!\!\!\!\!  = & \!\!\!\!\!\!
O^{J',1} O^{J-J',2} +
\sum_{m}\frac{g_2\,\sin^2(\pi m r)}
{{2} \sqrt{J}\,\pi^2 m^2} \, O^{J,12}_{m}~.
\eea
It can be checked that the new set of operators is orthogonal at order
$O(g_2)$. Actually the new basis introduced here (globally denoted
with $\hat\Omega$) is related to the one built from the BMN operators
in very simple way through the matrix $S_{\a\beta}$ appearing
in~\eq{ST}: $\hat\Omega = S^{-1/2} {\cal O}$. Eq.\eq{cob} is just the
first order expansion of this relation which automatically ensures the
orthogonality at all orders in $g_2$ of the $\hat\Omega$'s.

The main drawback of the choice Eq.\eq{cob} is that the $\Omega$'s are
neither operators with definite conformal dimension nor single trace
operators. Thus the dictionary involving the $\Omega$'s on the one
side does not have the mathematical simplicity of the one proposed
in~\cite{Janik:2002bd,Chu:2003qd,Georgiou:2003aa,Chu:2003ji} and on
the other does not follow the physical intuition discussed in
Section~\ref{bmndsect}. It is then natural to ask whether the
suggestive identification between the physical IIB strings and SYM
operators could be taken to hold also for $g_2\not= 0$, despite the
mixing discussed in the previous Section. This point of view has been
advocated in~\cite{DiVecchia:2003yp} mainly for the following reason:
all computations on the string side performed so far use the
oscillators $a_n$ and $b_n$ which are derived from the {\em local} (on
the world-sheet) actions Eqs.\eq{bos45}--\eq{act567} by solving the
resulting equations of motion on the whole {\em complex plane}. Thus
the predictions we can draw on the SYM side from these string
computations do not have general validity, but are expected to hold
only at the planar level and, even more restrictive, when only the
``local'' interactions, like that of Fig.\ref{fig-fterm}, are
considered. In fact, even if the states of the string Fock space are
orthogonal at the level of the sphere, there is no compelling reason
why the string $2$-point functions on the {\em torus} could not
develop non-diagonal terms. Thus, according to this point of view, the
string $1$-loop corrections should reproduce the $g_2^2$ terms of the
matrices $S$ and $T$ in Eq.\eq{ST}, by producing corrections to the
kinetic and mass term respectively in the effective action.
Unfortunately string amplitudes in the PP-wave background at the torus
level are very challenging and no explicit calculation is available in
the literature. However there are interesting indications that this
point of view is indeed correct, see~\cite{Huang:2002yt}, where an
approach inspired by quantum mechanical perturbation theory was used
to evaluate string loop amplitudes. Notice that the approach
of~\cite{Huang:2002yt} is similar in spirit to the one used in
Sections~\ref{mixsect} and~\ref{dif}.  Moreover, following the
arguments of~\cite{Balasubramanian:2001nh}, the identification between
the number of traces and the number of string states should be valid
as long as the SYM operators are not too ``big''. A simple
quantitative characterization of big operators can be derived by
realizing that overlap between single and double traces is of order
$\sqrt{J J' (J-J')}/N$. If this is not negligible in the planar
limit, then we are dealing with big operators. This shows that, even
if the BMN operators are made of an infinite number of fields, they
are never big since $\sqrt{J J' (J-J')}/N \sim g_2/\sqrt{J}\to 0$
(this is also the reason why the corrections in Eqs.\eq{h+}, \eq{chp}
and \eq{cob} are all suppresed by $1/\sqrt{J}$). Thus the usual rules
of AdS/CFT should apply: the single trace operators should correspond
to the elementary string states while the multi-trace operators should
be bound states and so they are not present in the spectrum of the
free string.  Moreover it was conjectured that the interaction with
the multi-trace operators is not captured by the usual string
interaction, but that, for this case, non-local interactions on the
string world-sheet are relevant~\cite{Aharony:2001pa}. Thus also the
terms of order $g_2$ in Eq.\eq{ST} can be interpreted in a more
conservative way without the need of a redefinition of the BMN
dictionary presented in Section~\ref{bmndsect}. It is in fact
suggestive that also on the field theory side the mixing is triggered
by ``non-local'' contractions (see Fig.\ref{nonpl-fig}). So it is more
natural to relate these contributions to the novel interaction
of~\cite{Aharony:2001pa} than to the usual string couplings of Section
\ref{3strsec}. 

\section{Comparing string and field theory results}\label{comparing}

\subsection{A proposal inspired by the AdS/CFT duality}\label{a}

The first proposal for the comparison of the string interaction with
the field theory was put forward in~\cite{Constable:2002hw}. This
proposal is motivated by the standard AdS/CFT dictionary between bulk
and boundary correlation functions~\cite{Gubser:1998bc}: since the
light-cone interaction vertex on the PP-wave in Eq.\eq{c3} can be
understood as the generating functional of the correlation functions
among string states, it is natural to put it in correspondence with
the correlation functions of the dual field theory operators. A
specific prescription was proposed in~\cite{Constable:2002hw} for the
leading order in $1/\mu$:
\be
\label{dual}
\frac{(\langle 1 | \otimes \langle 2| \otimes \langle 3|) ~|H_3
\rangle}{\sum_\rr \left(H_2^{\rr}\right)}
= C_{ijk}~, 
\ee
where $C_{ijk}$ is the coefficient appearing in the tree-level
correlator among three BMN operators of R-charge $J_i$. Again this
formula is very reminiscent of quantum mechanical perturbation theory
and thus it was conjectured to be valid when the energy difference in
the denominator is small in the $\mu\to\infty$ expansion ({\em i.e.}
for impurity preserving transitions). In this case, it is easy to
extract this SYM coupling by using the definition of the barred
operators introduced in~\eq{ob}
\be
\label{3pt}
\lim_{y\to\infty} \langle  \bar O_i(y)  O_j(x) O_k(0) \rangle
= {C_{ijk}} ~,
\ee
where $C_{ijk}$ depends on the R-charges $J_i$ and on the level $m$ of
the BMN operators.

From Eq.\eq{dual} it follows that the prefactor of the string
interaction vertex should reduce, or at least be proportional, to the
difference of the energies of the ingoing and outgoing states.
However, a careful analysis~\cite{Pankiewicz:2002gs} of the vertex
proposed in~\cite{Spradlin:2002ar} finally showed that this is not the
case (see \cite{Spradlin:2002rv} v3). The physical reason for this
failure is quite simple: if one insists in keeping a smooth flat space
limit $\mu\to 0$, then the vertex should obey for $\mu=0$ the
holomorphic factorization property and should appear as a product of
the left and right moving oscillators $\hat a_1^\dagger$ and $\hat
a_2^\dagger$, rather than the sum which appears in the free
Hamiltonian.
The alternative proposal for the supersymmetric completion of the
kinematical vertex Eq.\eq{nqh}, which has not a smooth flat space
limit, fits instead very well with the proposal Eq.\eq{dual}.  Indeed,
it is possible to show that the $3$-string couplings derived from the
vertex Eq.\eq{nqh} are in agreement with the free field theory
evaluation of the coefficients $C_{ijk}$ from the 3--point correlation
functions of single trace operators~\cite{DiVecchia:2003yp}. The
agreement holds, at first order in $1/\mu$, for arbitrary $3$-point
functions, regardless of the particular type of impurities, which can
be scalar, vector or fermionic fields, and of the type of correlators,
which can be impurity preserving or not. We focus here only on BMN
operators with fermionic impurities, referring
to~\cite{Kiem:2002xn,Huang:2002wf,Chu:2002pd} for the scalar and
to~\cite{DiVecchia:2003yp} for the vector impurities. The approach
here presented has been studied also from the string bit point of view
in~\cite{Zhou:2002mi}.
Due to the form of the prefactor in Eq.\eq{nqh}, we can focus only on
the kinematical part of the vertex Eq.\eq{kv2} and rewrite
Eq.\eq{dual} as:
\be
\langle i | \langle j | \langle k |V \rangle =
{C_{ijk}}/{C^{(0)}_{ijk}} ~,
\label{come}
\ee
where $C^{(0)}_{ijk}=\sqrt{J_1J_2J_3}/N$ is the combinatorial factor
of the Green function among three vacuum operators Eq.\eq{bpsdic}.  This
factor ensures the same normalization for the two sides of Eq.\eq{come},
since the string overlap ${}_{123}\langle v |V\rangle$ is set equal to
one.

On the field theory side, the evaluation of the $3$-point function of
BMN operators with fermionic impurities can be reduced, by means of the
identities Eqs.\eq{lac}, to the same combinatorics found in the
case of scalar impurities. One has therefore, for double--impurity operators:
\be
\lim_{x_3\to\infty} \langle \bar O^{J_3}_{\a \b,n}(x_3)  
O^{J_2}_{\a \b,m}(x_2) O^{J_1}(x_1) \rangle
= - \frac{J_2}{N}\sqrt{\frac{J_1J_2}{J_3}}\frac{\sin^2(\pi n
  y)}{\pi^2(m-ny)^2} ~,
\label{fri-2}
\ee
with $J_3=J_1+J_2$, $y=J_2/J_3$ and $\a\not= \b$. The only difference with the
correlators containing scalar impurities is that one can get some
extra minus signs, due to the anticommuting nature of the fermions.
This difference becomes important for correlators among operators
containing spinors of the same flavour
\be
\langle \bar O^{J_3}_{\a \a,n} O^{J_2}_{\a\a,m} O^{J_1}_{vac} \rangle
= \langle \bar O^{J_3}_{\a\b,n} O^{J_2}_{\a\b,m} O^{J_1}_{vac}\rangle
- \langle \bar O^{J_3}_{\a\b,-n} O^{J_2}_{\a\b,m} O^{J_1}_{vac}\rangle ~,
\label{fing}
\ee
On the string theory side, according to the dictionary Eq.\eq{fermdic},
one has to compute the amplitude~\eq{come} with the external states
$\bra{\a_1,\a_2,\a_3} \theta_{n(3)}^{1 a}\theta_{n(3)}^{2 b}
\theta_{m(2)}^{1 a}\theta_{m(2)}^{2 b}$
\be
A^{ab}_f = \label{aijf}
=  -\frac{1}{4}\left[(Q^{23}_{mn})^2+ (Q^{32}_{nm})^2
- 2 Q^{23}_{mn}Q^{32}_{nm}\right] ~,
\ee
and to extract the first term in the $\mu\to\infty$
limit. This can be done by using Eq.\eq{defQ1} and the leading
term~\cite{Huang:2002wf} in the expansion of the Neumann
matrices\footnote{A detailed study of the subleading terms in the
  $1/\mu$ expansion can be found in~\cite{He:2002zu}. See
  also~\cite{Lucietti:2003ki}, were methods of complex analysis are
  used to tackle a closely related problem.}
\bea
N^{32}_{nm}&\sim& \frac{2}{\pi}\frac{ny^{3/2}\sin(\pi
  ny)}{m^2-n^2y^2}~,~~~{\rm with}~~n,m>0
\label{nlim}
\\
U_{(i)}&\sim& \frac{n}{2\mu\a_i}~, \quad\quad 
U_{(3)}\sim -\frac {2\mu\a_3}{n}~.
\label{ulim}
\eea 
By using these formulae, one can check that the string Eq.\eq{aijf}
and the gauge theory Eq.\eq{fri-2} results agree. The string amplitude
related to Eq.\eq{fing} is $A^{aa}_f = Q^{23}_{mn}Q^{32}_{nm}$. Notice
that the quadratic terms have disappeared, because the fermionic
nature of the oscillators implies $(\theta_n^a)^2=0$. Anyway also in
this case, by using Eq.\eq{nlim} and Eq.\eq{ulim} one recovers exactly the
gauge theory results Eq.\eq{fing}.

So far we considered only extremal correlators, {\em i.e.} correlators
where in the Born approximation there are no propagators connecting
the operators $O^{J_2}$ and $O^{J_1}$. In this case the field theory
computation inherits some of the properties of the $2$--point
functions. One may wonder whether Eq.\eq{come} is of more general
validity as it is suggested by the analysis
of~\cite{Dobashi:2002ar,Yoneya:2003mu}.  This possibility is motivated 
by considering the Penrose limit of the AdS/CFT bulk--to--boundary
formula~\cite{Gubser:1998bc}.
Actually the simplest possible check supports this possibility. In
fact the classical contribution to the non--extremal correlators
vanishes in the BMN limit faster than ${C^{(0)}_{ijk}}$, so that the
ratio Eq.\eq{come} is zero. Correspondingly $N^{ij}_{nm}$ with
$i,j=1,2$ tend to zero when $\mu\to \infty$ and thus all non--extremal
amplitudes are vanishing at leading order in $\lambda'$ both on the
string and on the field theory side.
%%%%%%%%%%%%%%%%%%%%%%%%%%%%%%%%%%%%%%%%%%%%%%%%%%%%%%%%%%%%%%

Notice that the subleading contributions to the Neumann matrices
corresponding to these correlators start with a term of the order of
$\sqrt{\lambda'}=\sqrt{\lambda}/J$~\cite{Klebanov:2002mp}.  It is
interesting to observe that this particular $J$ dependence is already
captured by the free field theory computation of the non--extremal
correlators. This can be seen by considering the scaling with $N$
and $J$ of the simplest non--extremal correlator between supergravity
states
\be
\frac{1}{N_o^{J+1}\sqrt{J}} 
\langle \Ttr[\bar Z^J ](0) \Ttr[\phi Z^{J_1}](x_1) \Ttr[\bar\phi
Z^{J_2}](x_2) \rangle 
\sim \frac{\sqrt{J}}{N}~.
\label{non-extr}
\ee 
From Eq.\eq{non-extr} it follows that the $J$ dependence in the
r.h.s.~of Eq.\eq{come} for these correlators is $1/\sqrt{J_1 J_2}$,
which matches with the $\alpha_i$'s dependence of the corresponding
Neumann matrix element in the $\mu\to\infty$ limit $N^{12}_{00}\sim
1/\mu\sqrt{\alpha_1\alpha_2}=\sqrt{\lambda}/\sqrt{J_1J_2}$~\cite{He:2002zu},
upon using the relation Eq.\eq{rela44a} between $J$ and
$\alpha=\alpha_1+\alpha_2$. Moreover, the peculiar dependence on the
square root of the coupling $\sqrt{\lambda}$ is reminiscent of the
behaviour of other dynamical observables evaluated in the strong
coupling regime of field theory by means of AdS/CFT duality, as the
vacuum expectation value of the Wilson loop~\cite{Semenoff:2002kk}.
Notice that the string theory prediction implies a non-trivial
dependence on the coupling also for the $3$-point non-extremal
correlators between supergravity states. This is contrary to the conjecture
based on the supergravity analysis of~\cite{Lee:1998bx} and to the
field theory arguments that support it, see References
in~\cite{D'Hoker:2002aw}.  From this point of view, the most natural
expectation is that the string result is recovered after a
resummation~\cite{Klebanov:2002mp} of the planar SYM diagrams.

\begin{figure}[htb]
\begin{center}
\begin{picture}(0,0) %magnification factor 
\includegraphics{cordeint.ps}%
\end{picture}%
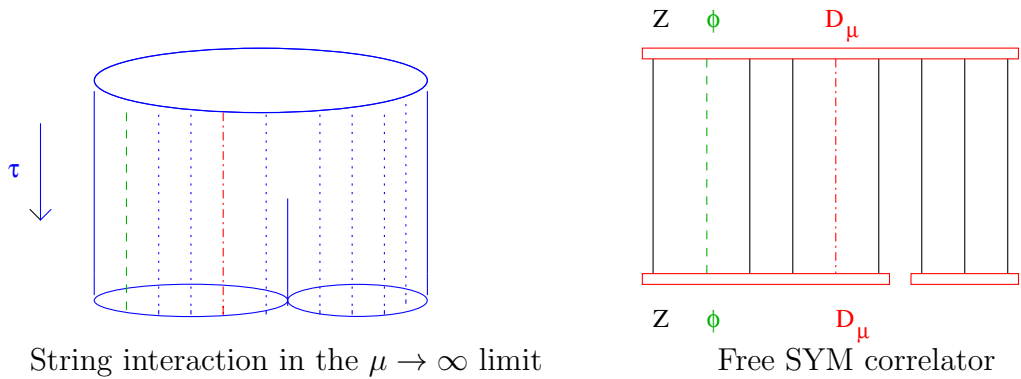
\end{center}
\caption{\label{fig5} In the $\mu\to \infty$ limit the various ``bits'' of the string can not interact. Pictorially this exactly matches the field theory behaviour in the $\lambda' \to 0$ limit.}
\end{figure}

One positive feature of the approach in Eq.\eq{dual} for the
comparison with the field theory is that there is a clear pictorial
understanding of the duality between string and gauge
theory~\cite{Constable:2002hw}. In fact in the $\mu\to \infty$ limit
the kinetic term of the free string world--sheet Lagrangian
Eqs.\eq{bos45} and \eq{act567} can be neglected with respect to the
mass term.  If one discretises the string in $J$ bits, these will be
then represented by a bunch of independent harmonic oscillators. The
combinatorics obtained by imposing the smoothness of the world--sheet
during the interaction is exactly reproduced by the free correlators
of the gauge theory BMN operators, in which each elementary field
represents a bit of the string, see Fig.\ref{fig5}.

One of the open questions in this approach is to understand how to
handle the space-time dependence of the field theory correlators. This
problem becomes particularly important when considering the $1/\mu$
corrections. In fact, at subleading orders in $\lambda'$, Eq.~\eq{3pt}
depends non-trivially on $x$. In order to have the well defined
scaling behaviour of Eq.\eq{3pt} at one loop level, it is necessary to
take into account the mixing between single and double trace
operators~\cite{Bianchi:2002rw,Beisert:2002bb,Constable:2002vq}. The
puzzling point is that this mixing can modify the tree-level value
of $C_{ijk}$ and spoil the agreement with the string computations
described here. 
A closer inspection reveals that this phenomenon is just the PP-wave
version of the usual peculiarity of extremal correlators, where the
contribution of the multi-trace operator is enhanced and becomes of
the same order as the leading terms. In the standard AdS/CFT duality,
the simplest comparison between the supergravity and field theory
results is done without advocating any
mixing~\cite{Liu:1999kg,D'Hoker:1999ea} and agreement is found. In the
PP-wave case, the problem is more severe just because we deal with
non-protected operators. It is interesting to notice that exactly for
these correlators the requirement of planarity does {\em not} ensure
the ``locality'' of the interaction, and diagrams connecting non
contiguous impurities are as important as the others. This may be the
reason for the problems in comparing the extremal correlators with the
string results.

\subsection{A proposal inspired by Eq.\eq{rf}}\label{dif}

A different approach to the comparison between string interactions and
gauge theory data was proposed in~\cite{Gross:2002mh}. This approach
was inspired by the string bit proposal of~\cite{Verlinde:2002ig}, in
which it was suggested to interpret the relation between the
dilatation operator and the light-cone Hamiltonian~\eq{rf} ($\Delta-J
= H/\mu$) as an {\em exact} operatorial equation, including the
non-linear corrections on the string theory side. This means that in
the r.h.s.~of Eq.\eq{rela45a} we should include, at first order in
$g_2$, also $\ket{H_3}$, which is the first correction to $H$ (see the
discussion after Eq.\eq{h3}).  Since these corrections are non-linear,
the full Hamiltonian acts also on states that live in the $n^{\rm th}$
tensor product of the free string Hilbert space, see Eq.\eq{c3}. Thus,
also the SYM operator $\Delta-J$ should act in a bigger space than the
one considered in Section~\ref{bmndsect}. At the light of the
discussion in Section~\ref{mixsect}, the most natural proposal is to
identify the states living in the tensor product of two or more string
Hilbert spaces with multi-trace operators. This means that, at first
order in $g_2$, we should work with the sets of SYM operators ${\cal
  O}_\a$ introduced just before Eq.\eq{ST}. On the one side this
mapping may appear ``obvious'', but on the other side we should notice
that it represents a departure from the common interpretation of the
AdS/CFT duality. In fact, usually multi-trace operators are mapped to
bound states on the string side (see point $3$ in Section~\ref{basic})
and not to multiparticle states. The PP-wave limit might be particular
because the harmonic potential confines the excitations around
$X^I=0$, but notice that there is still the possibility to separate
the constituent particles along the $x^+$ direction.

There are basically two different ways to check this proposal. A first
possibility is to diagonalize the operators both on the string and on
the gauge theory side and then compare their eigenvalues. Conceptually
this is the most straightforward approach, but it is clearly quite
challenging from the computational point of view. We already discussed
the diagonalization procedure on the SYM side in
Section~\ref{mixsect}, where the primary operators~\eq{chp} were
introduced. In fact, from Eq.\eq{dg2} we can directly read the
eigenvalues of $\Delta-J$. Thus we now need to extract the $g_2^2$
corrections to the energy of the string states. In principle this
requires the computation of the string $2$-point function on the
torus, since we know that different powers of $g_2$ are related to
diagrams of different topology. As we already said, full fledged
string computations of this type are not available for the PP-wave
background. Thus we will follow the same approach used on the gauge
theory side to derive Eq.\eq{dg2}: we will apply quantum mechanical
perturbation theory, by interpreting the corrections to the free
Hamiltonian as perturbation terms and $g_2$ as the small parameter
governing the perturbative expansion.  Since we know that the first
non trivial correction to the energy is of order $g_2^2$, we need to
include also the correction to the classical string Hamiltonian up to
the same order. Thus in principle we should extend the analysis of the
string interaction done in Section~\ref{3strsec}. However, by
following~\cite{Roiban:2002xr}, we can extract the relavant part of
$g_2^2$ Hamiltonian by using the supersymmetry algebra~\eq{sal1} which
implies that $Q^2 \sim H$. In particular the $g_2^2$ correction to $H$
relevant for the matrix element of single string state can be derived
by using the $g_2$ correction to the supercharges presented in
Section~\ref{3strsec}. Thus the energy shift for a string state like
the one of Eq.\eq{bosdic2} is
\begin{equation}\label{soe}
\delta E_n^{(2)}= g_2^2 \sum_{s} \frac{1}{2} 
\frac{\left|\langle n|H_3| s \rangle
  \right|^2}{E_n^{(0)}-E_{s}^{(0)}} 
+\frac{1}{8}\sum_{\dot{a},s}\left|\langle n|
Q_{3\,\dot{a}}|s\rangle\right|^2~,
\end{equation}
where we should sum over all states $\ket{s}$ satisfying the
level-matching condition. The first term comes from the second order
perturbation theory and because of this has an overall factor of
$1/2$, while the second term is the first order contribution coming
from the $g_2^2$ correction of the Hamiltonian. The factor of $1/8$ is
due to the trace over the spinor indices ({\em i.e.} the sum over
$\dot{a}$). There are various explicit computations in the
literature~\cite{Roiban:2002xr,Gomis:2003kj,Pankiewicz:2003kj} that
use~\eq{soe} with bosonic external states in different representations
of $SO(4)\times SO(4)$. The result of these computations is that
Eq.\eq{dg2} is always reproduced, when the sum over $m$ is restricted
to the $2$ impurity states, so that all the amplitudes entering
in~\eq{soe} are impurity preserving. 
This approximation is similar to the one discussed on the gauge theory
side (see the discussion after~\eq{dg2}), however a physical
justification of this truncation is not available.

A second possible test of this approach is to find a dictionary
between string and gauge theory Hilbert spaces so that all elements of
the ``matricial'' relation $\Delta-J = H/\mu$ can be compared.
Unfortunately the simple BMN dictionary discussed in
Section~\ref{bmndsect} is not the right basis for the proposal here
discussed and we need to generalize the BMN dictionary along the lines
explained in Section~\ref{bmnrev}. Even if it is difficult to find a
clear physical principle that can substitute the idea resumed in
Figure~\ref{op-st}, it is interesting to see that the
redefinition~\eq{cob} yields results that are consistent with those
extracted from the string vertex~\eq{h}. This approach uses the matrix
$S$ appearing in~\eq{ST} to define the basis of SYM operators dual to
string states. This is sometimes called string basis because the
quantum corrections to the dilatation operators match the
results derived from the string vertex~\eq{h}
\beq
S^{-1/2} T_{\a\beta} S^{-1/2} = \langle \a,\beta \ket{H_3}^I~.
\eeq
This relation was checked in a number of
cases~\cite{Gomis:2002wi,Gomis:2003kj,Gomis:2003kb,Georgiou:2003kt} at
the leading order in $\lambda'$ and for various bosonic
impurities. The first checks at subleading orders in the $1/\mu$
expansion show however a disagreement~\cite{Spradlin:2003bw}.

\subsection{Open problems}

Let us conclude by recalling the main open problems in the comparison
between interacting string theory on the PP-wave~\eq{pwave-back} and
the BMN subsector of ${\cal N}=4$ SYM.

%\begin{itemize}

-- %\item 
We saw that the light-cone construction of the $3$-string vertex
  is not completely unambiguous and that the $\zet_2$ parity can be
  implemented in different ways. It would be interesting to have a
  more thorough understanding of this freedom. This point is strictly
  related to the $\mu\to 0$ limit. From the
  point of view of the duality the smoothness of this limit appears
  very puzzling. In fact this would 
  imply that the (interacting) IIB string theory in 10 dimensional
  flat space can be completely captured by studying  the anomalous
  dimensions of BMN operators.
  
-- %\item 
An important point that deserves further study is the operator mixing.
It is sometimes said that this is a peculiar feature of the PP-wave,
where it would be impossible to separate single and multi-trace
operators. However, the discussion in Section~\ref{mixsect} shows that
this is a general phenomenon. As is shown 
in~\cite{Bianchi:2002rw,Arutyunov:2002rs},
for short operators the mixing is suppressed by a factor of $1/N$
exactly as in the PP-wave case it is suppressed by a factor of
$g_2$~\cite{Beisert:2002bb,Constable:2002vq}. In the light of this
observation it would important to clarify the role of the mixing in the
string/gauge theory dictionary, since the same problem will also appear
in the full AdS/CFT correspondence.
  
-- %\item 
  Let us now focus on the approach outlined in Section~\ref{a}. Here the main
  open problem is to see how to extend the validity of Eq.\eq{come} also
  at the subleading orders in the $1/\mu$ expansion. It is possible to
  study this issue in two different situations. In the case of
  impurity preserving amplitudes, the expansion on the string side is
  simple, as it involves integer powers of $\lambda'$. However, on
  the SYM side, the $3$-point functions have a non--canonical space-time
  dependence~\cite{Constable:2002vq}. Moreover, in this case a better
  understanding is necessary of the role
  played in the duality by the SYM diagrams connecting non-contiguous
  impurities. Another possibility is to study
  non-extremal correlators, where the gauge theory side of the
  computation is simpler, at least conceptually. In this case the
  string predictions are surprising. First the large $\mu$ expansion
  of the Neumann matrices yields non-integer powers of $\lambda'$,
  which probably means that the limit Eq.\eq{bmnlimi85} can not be taken
  order by order in perturbation theory any more. Then even the
  simple computation in Eq.\eq{non-extr} presents an unexpected
  behaviour. It is obviously very important to analyze
  this amplitude more carefully.
  
-- %\item 
As for the approach outlined in Section~\ref{dif}, it is first necessary
  to obtain a better understanding of the impurity non-preserving
  amplitudes and the truncation used to derive Eq.\eq{soe}. In the
  impurity preserving sector the proposal is in principle
  self-consistent for all values of $\lambda'$. However,  as was already said,
  the first next-to-leading computation~\cite{Spradlin:2003bw} 
  shows a disagreement.
  It would be interesting to see whether this problem persists also at
  the level of eigenvalues, or it is due to the proposed
  dictionary~\eq{cob}. Finally, all computations done so far within
  this approach deal with bosonic impurities. As a result, only a
  small part of the prefactor~\eq{h} has been checked. To overcome
  this limitation, it is clearly important to take fermionic
  impurities into account. It would be particularly interesting to see
  whether the non-diagonal parts of $v_{IJ}$ can be derived from
  field theory computations.

%\end{itemize}

%%%%%%%%%%%%%%
 
{\bf Acknowledgement} We would like to thank our collaborators P.~Di
Vecchia, J.~L. Petersen, and M.~Petrini for many discussions on the
topics of this report. We thank C.~Kristjansen for careful reading of the
manuscript. This report was presented at the EC-RTN workshop
``The quantum structure of spacetime'' held September 2003 in Copenhagen.
We would like to thank the organizers and participants of
that conference for the lively atmosphere and the many interesting
questions and discussions.
Our work has been supported is supported by EU Marie
Curie Fellowships and by EU RTN contracts HPRN-CT-2000-00122 and
HPRN-CT-2000-00131.

\section*{References}
%\bibliographystyle{h-physrev3}
%\bibliography{bibpp}

\end{document}